\documentclass[lettersize,journal]{IEEEtran}
\usepackage{amsmath,epsfig,lipsum,booktabs,xcolor,mathtools}
\usepackage{hyperref}
\usepackage{amsfonts}
\usepackage{algorithmic}
\usepackage{algorithm}
\usepackage{epsfig}
\usepackage{times}
\usepackage[caption=false,font=normalsize,labelfont=sf,textfont=sf]{subfig}
\usepackage{textcomp}
\usepackage{stfloats}
\usepackage{url}
\usepackage{verbatim}
\usepackage{graphicx}
\usepackage{cite}
\usepackage{xcolor}
\usepackage{multicol}
\usepackage{cuted}
\usepackage{orcidlink}
\usepackage{tikz}
\usepackage{balance}
\usepackage{rotating}
\usepackage{graphicx}
\usepackage{forest}
\usepackage{multirow}
\newcommand{\rev}[1]{{\color{black}#1}}
\newcommand{\revb}[1]{{\color{black}#1}}
\def\BibTeX{{\rm B\kern-.05em{\sc i\kern-.025em b}\kern-.08em
    T\kern-.1667em\lower.7ex\hbox{E}\kern-.125emX}}

\usepackage[utf8]{inputenc}
\usepackage{enumitem}

\begin{document}
\title{\rev{A Survey of Deep Learning Video Super-Resolution}}
\author{Arbind Agrahari Baniya\orcidlink{0000-0002-9359-6506},  
Tsz-Kwan Lee\orcidlink{0000-0003-4176-2215}, Peter W. Eklund\orcidlink{0000-0003-2313-8603}, and Sunil Aryal\orcidlink{0000-0002-6639-6824}}
\maketitle
\vspace{-3em}
\begin{abstract}
Video super-resolution (VSR) is a prominent research topic in low-level computer vision, where deep learning technologies have played a significant role. The rapid progress in deep learning and its applications in VSR has led to a proliferation of tools and techniques in the literature. However, the usage of these methods is often not adequately explained, and decisions are primarily driven by quantitative improvements. Given the significance of VSR's potential influence across multiple domains, it is imperative to conduct a comprehensive analysis of the elements and deep learning methodologies employed in VSR research. This methodical analysis will facilitate the informed development of models tailored to specific application needs. In this paper, we present an \rev{overarching} overview of deep learning-based video super-resolution models, investigating each component and discussing its implications. Furthermore, we provide a synopsis of key components and technologies employed by state-of-the-art and earlier VSR models. By elucidating the underlying methodologies and categorising them systematically, we identified trends, requirements, and challenges in the domain. As a first-of-its-kind \rev{survey} of deep learning-based VSR models, this work also establishes a multi-level taxonomy to guide current and future VSR research, enhancing the maturation and interpretation of VSR practices for various practical applications.
\end{abstract}



\maketitle
\vspace{-1em}
\section{Introduction}
An increase in the consumption of video-based multimedia in recent years can be attributed to advances in video capture technology, transmission networks, and rendering devices. These advances have led to an increased demand for higher-quality video signals. Quality can be defined from two standpoints, the quality of service (QoS) and quality of experience (QoE). From a QoS perspective, higher quality refers to a video bitstream with a higher bitrate, larger spatial resolution, and/or higher temporal resolution (more frames per second). While from a QoE perspective, higher quality is subjective and can be difficult to quantify as it aligns with a more pleasing perception, a judgement that varies greatly among individuals. It has been established that enhancements along the spatiotemporal dimensions of video signals result in an improved quality that positively correlates to an improvement in QoS and is, in turn, linked to perceived improvements in QoE~\cite{zhang2020improving}. The enhanced resolution improves various aspects of video quality and user experience. As a result, video super-resolution (VSR) models are widely being developed~\cite{purohit2020mixed,sun2020attention}, as the process of generating high-resolution (HR) video output with improved quality from a given low-resolution (LR) video input. Assuming a high-resolution video has undergone the following operation:

\vspace{-0.8em}
\begin{small}
\begin{equation}
     LR = (HR*k){\downarrow}_{d}+ns 
     \label{eqn:vsr_def}
\end{equation}
\end{small}

where $LR$ is the low-resolution video when each frame of high-resolution video $HR$ is convoluted with a blur or cubic kernel $k$ followed by downsampling operation $d$ and addition of noise $ns$, super-resolution of $LR$ is then a task of estimating the kernel, downsampling operation, and the noise such that $HR$ video can be obtained inversely from $LR$ video. As implied by eqn.~(\ref{eqn:vsr_def}), VSR is an ill-posed inverse problem that is considered an open research area in the low-level computer vision domain. VSR has mostly been treated as an extension of single-image super-resolution (SISR) and multi-image super-resolution (MISR). However, unlike SISR and MISR, modelling the tasks in VSR is challenging due to the need to aggregate highly correlated but misaligned frames in a given video sequence~\cite{DCRnet,cheng2013classification}. Adopting the approaches used in conventional SISR and MISR designed for an image directly to video-based super-resolution may fail to capture the temporal reliance between video frames~\cite{lee2021dynavsr,kramer2011local}. Therefore, recent studies have adopted learning-based approaches that exploit spatiotemporal features present in an LR video to super-resolve it to HR video~\cite{RSDN,TGA,liu2021efficient,EDVR,li2022video,BasicVSR}. 

Traditionally, upsampling algorithms such as back-projection methods~\cite{cohen2000polyphase} and least mean squares (LMS) based Kalman filter methods~\cite{costa2007statistical} have been used to interpolate pixels in a video frame or a single image. These methods rely on deterministic functions to map LR input to HR output. However, the deterministic nature of traditional methods limits their capacity to generalise well over different videos, and the inverse function obtained by traditional methods does not capture the non-linear complexity of the transformation that maps HR to LR video. Consequently, deep learning VSR models have garnered significant attention in recent times owing to their stochastic and data-centric characteristics, enabling effective generalisation across diverse video inputs. Moreover, these models possess the capacity to learn non-linear functions for mapping LR videos to their HR counterparts. Learning-based methods for VSR typically comprise feature extraction, alignment, fusion, reconstruction, and up-sampling as fundamental steps in the super-resolution process. The extraction of relevant features from accurately aligned frames and their subsequent fusion is of utmost importance in such models~\cite{BasicVSR,RBPN,fang2022high}. In this paper, we thoroughly investigate each component of deep learning-based VSR models. To date, there has only been one work published in this space~\cite{liu2022video}; however, the complexity in the VSR domain is significantly diluted with a single-layer taxonomy focusing only on the alignment step. Several other components within VSR are remarkably diverse and thus contribute to increasingly varied outcomes that are often difficult to interpret and explain. This paper aims to bridge these gaps by:
\begin{itemize}
    \item \rev{developing a novel taxonomy and extensive listing of approaches and trends within individual VSR components;}  
    \item providing a thorough methodological review of deep learning in the context of video super-resolution;
    \item providing a comprehensive overview of VSR literature, current trends, applications, and challenges;    
    \item making the VSR models and their respective performances more explainable;
    \item and providing future VSR works with guidelines based on the prospective requirements and gaps.
\end{itemize}
\vspace{-1em}
\section{Background}

\subsection{Image Super-Resolution (ISR)}
Single image super-resolution (SISR) is a technique used to enhance the resolution of a single image by increasing the number of pixels in the image\rev{~\cite{CHEN2022124}}. The goal of SISR is to generate a high-resolution image from a low-resolution counterpart by interpolating missing pixels and adding high-frequency details. At the same time, multi-image super-resolution (MISR) is a technique used to enhance the resolution of single or multiple images by combining the infusion of pixels from multiple images\rev{~\cite{LEPCHA2023230}}. The earliest methods for image super-resolution were based on interpolation, such as nearest-neighbour~\cite{efros1999texture} and bicubic interpolation. These methods are simple to implement and computationally efficient but could not exploit the high-frequency information present in the images, leading to artefacts and a lack of details. Example-based ISR methods were then proposed~\cite{kim2008example}, such as self-exemplar and sparse-coding-based methods. These methods used external examples to guide the interpolation process, improving the quality of the generated HR images. However, they are limited by the quality and availability of the examples. Similarly, image prior-based super-resolution techniques exploited prior knowledge about the regularities and structures found in high-resolution images to super-resolve low-resolution inputs~\cite{sun2008image,tai2010super}. The assumptions and restrictive adaptability of priors-based approaches result in limited applicability and generalisability while involving computationally complex optimisation algorithms.

\rev{Deep learning-based ISR methods, such as convolutional neural networks (CNNs), recurrent neural networks (RNNs), transformers and generative adversarial networks (GANs), have been gaining popularity\rev{~\cite{wang2020deep,SU202246}} because of their ability to learn complex and non-linear mappings from LR to HR images and generate high-quality HR outputs by exploiting the low-frequency information present in the input~\cite{dong2014learning}. The success of these approaches is significantly attributed to the deep neural networks' ability to automatically extract features of interest and the hierarchical representational ability of the non-linear complex patterns required to be restored for super-resolution. More recently, efficiency concerning deep learning is leading ISR models to mitigate complex networks with high computational and memory demands. Lightweight CNN and transformer backbones are being used to significantly reduce memory usage compared to traditional transformers~\cite{lu2022transformer}. However, the trade-off between reconstruction performance and efficiency is to be balanced better. An example is the usage of hierarchical dense residual blocks incorporated without significantly increasing computational overheads~\cite{jiang2020hierarchical}. Non-local operations and sparse representations are also being utilised to improve both quantitative metrics and visual quality~\cite{mei2021image}. Other alternatives tackling the computational demands in transformers incorporate shift convolution and group-wise multi-scale self-attention modules, offering superior performance with significantly reduced computational complexity~\cite{zhang2022efficient}. Nevertheless, overcoming computational challenges while maintaining or surpassing state-of-the-art performance remains a focused trend in current ISR research~\cite{LEPCHA2023230}.}

\vspace{-1em}
\subsection{From Image to Video Super-resolution}
The fundamental difference between images and video is that video comprises multiple frames over an added temporal dimension. This temporal dimension of video adds additional complexity to the super-resolution task, as it requires aligning and fusing multiple temporally dispersed frames to generate a high-resolution video. Extending the target-resolving subject from image to video signals, super-resolution approaches used in conventional ISR to VSR fail to capture the unique temporal information present in videos. VSR aims to adopt several temporally correlated low-resolution frames within a video sequence to super-resolve the frame series. Considering spatial and temporal dimensions across multiple input frames induces VSR as a highly non-linear multi-dimensional problem. \rev{VSR serves online and offline application contexts addressing three primary objectives: enhancing QoS, improving QoE, and assisting computer vision systems. It elevates the sharpness and detail of low-resolution videos in platforms like video streaming and multimedia communication, enriches the visual experience in entertainment and gaming, and aids accurate analysis and recognition in computer vision systems, essential for surveillance, autonomous vehicles, and robotics.}

Attempts have been made to translate the ISR problem definition and solutions to VSR. For instance, many early VSR methods directly applied ISR methods, such as interpolation-based, example-based~\cite{watanabe2008fast} and video prior-based~\cite{kong2006video} methods, to video frames without considering the temporal dimension of the video. However, these methods led to temporal inconsistencies and a lack of detail with a smoothing effect in the generated HR video~\cite{7001251}. Improving on this, VSR methods began incorporating additional techniques such as motion estimation and compensation (MEMC) and temporal fusion in addition to the conventional ISR methods to exploit the spatiotemporal correlation present in the video. This improved the quality of the generated HR video, resulting in a more realistic and natural output. Examples of such methods include traditional methods like Kalman filter-based methods~\cite{790425,newland2007modified} and adaptive filtering methods~\cite{748893}, and more recently, deep learning-based methods~\cite{Liao_2015_ICCV}.

Deep learning-based methods have been widely adopted in VSR due to their effectiveness in automatically extracting and learning desired features by leveraging the spatiotemporal information present in the video. The use of deep learning has enabled the development of end-to-end VSR models that can learn the mapping between LR and HR videos in a data-driven manner. Earlier deep-learning models tried to learn the mapping between LR and HR frames by extracting residual features using LR frame(s) and motion details~\cite{RBPN}. These early models still lacked the sequential modelling ability desired to learn the true nature of videos. \rev{As a result, there is a growing interest in developing VSR models for exploiting temporal dependencies between video frames and capturing long-term spatiotemporal patterns across the time domain.} One popular approach is to use recurrent neural networks (RNNs) with memory-preserving techniques such as residual blocks with skip connections and long-short term memory (LSTM) to model the temporal dynamics of the video. RNN-based VSR models effectively learn the underlying temporal patterns and capture the long-term dependencies between frames, resulting in temporally consistent reconstructed HR videos~\cite{RRN, BasicVSR, baniya2023online}. Another approach is to use CNNs with 3D convolutions to capture both spatial and temporal features in the video~\cite{DUF,3dsrnet}. These models can effectively learn the spatiotemporal patterns present in the video and provide a more accurate mapping between LR and HR frames. However, the extent of temporal context that 3D CNN-based models can learn is limited to a fixed temporal window, unlike the global information propagation in RNNs.

\begin{figure*}[ht]
 \vspace{-3em}

    \centering
        \includegraphics[width=0.8\textwidth]{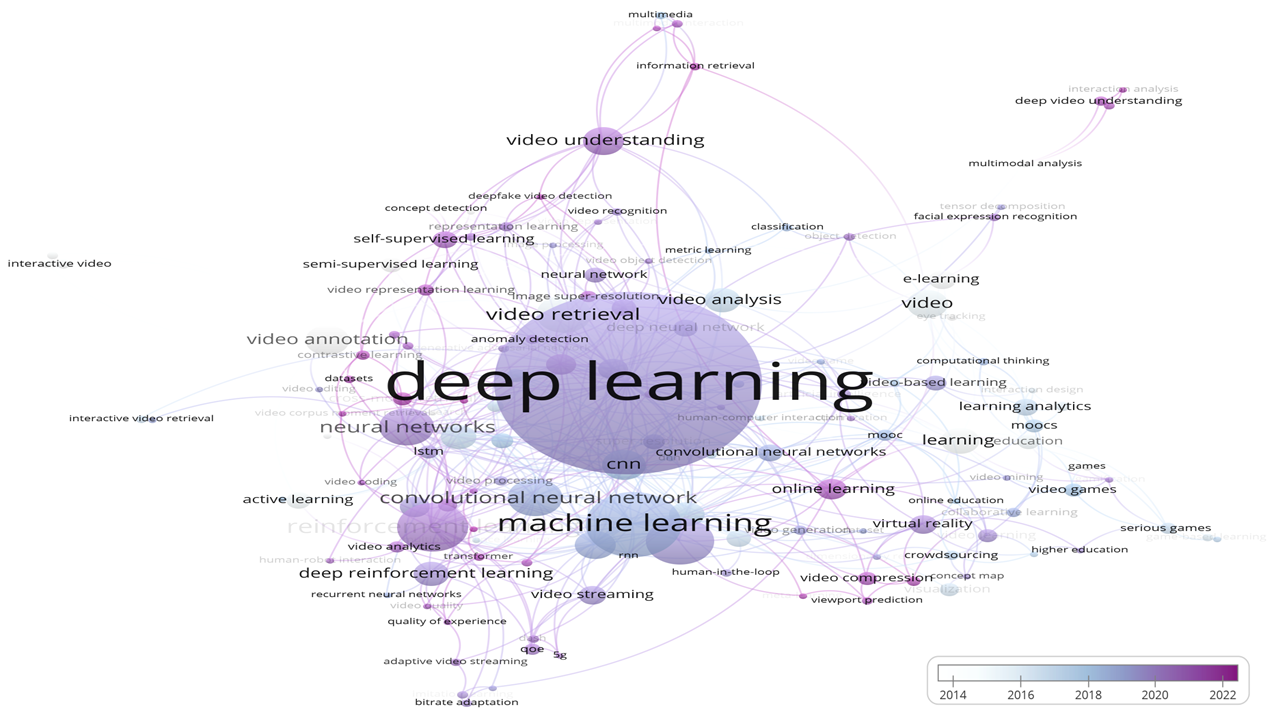}
         \vspace{-1em}

        \caption{\rev{Publication keyword cloud in last two decades with circle size representing number of occurrences and colour highlighting chronological occurrence.}}
        \label{fig:word_map}
         \vspace{-1.5em}

\end{figure*}

\begin{figure}[ht]
 \vspace{-1em}
    \centering
        \includegraphics[width=0.48\textwidth]{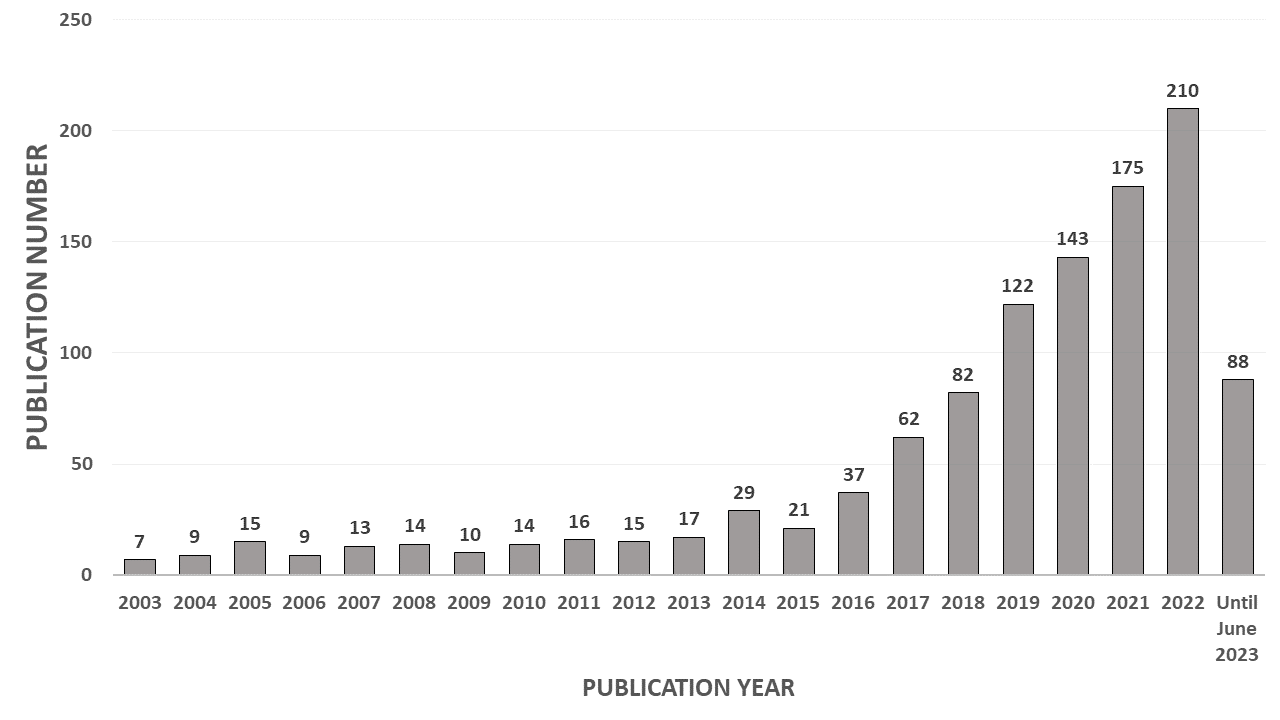}
         \vspace{-1em}
        \caption{The number of publications in deep learning for video super-resolution in the last two decades by year.}
        \label{fig:publications}
         \vspace{-1.5em}

\end{figure}

Furthermore, recent advancements in transformers using attention mechanisms have led to the development of attention-based VSR models. These transformers selectively focus on relevant frames and regions within the video, improving the reconstruction quality by attending to the most informative aspects of the video~\cite{TGA, li2020learning}. These attention-based VSR models effectively capture the spatiotemporal patterns in the video and have significantly advanced the state-of-the-art by improving the reconstruction quality of LR videos. However, as the field progresses, increasingly employing learning technologies to model the task of VSR, there are many unexplored sequential modelling advances for further investigation. To analyse the trend of work being done in VSR using deep learning, we explore the two popular databases, namely IEEE Xplore, with a search term \revb{(("All Metadata": 'video super-resolution') AND ("All Metadata": 'deep learning'))} and ACM Digital Library with a search term ([Title: 'video super-resolution'] AND [Title: 'deep learning']) for works published in the past two decades. We collect a total of 1,108 published works and plot a yearly trend as shown in Fig.~\ref{fig:publications}. The figure demonstrates a steadily increasing trend, particularly escalating in the last decade.  We further extract the keywords from these publications and plot a keyword map showcasing the occurrences of keywords and relationships as shown in Fig.~\ref{fig:word_map}. Evident here (among other things) is the greater use and frequency of "deep learning" and "convolutional neural networks" reinforcing more recent and popular key terms.

\vspace{-1em}
\subsection{Super-resolution of Emerging Video Formats}
\subsubsection{Omnidirectional Video Super-resolution}
Omnidirectional videos, also known as {360\textdegree} videos, capture the complete {360\textdegree} horizontal view of the surrounding environment and provide an immersive viewing experience. These videos are primarily used in virtual reality (VR) and augmented reality (AR) to provide immersive experiences as part of an extended reality. Omnidirectional video super-resolution aims to enhance the resolution of these videos to improve the visual quality and details it to improve the immersivity~\cite{fakour2018360}. The unique characteristics of omnidirectional videos, such as the equirectangular projection, pose challenges for traditional video super-resolution methods. The spatially varying distortion in the equirectangular projection and the need to handle {360\textdegree} coverage requires specialised techniques for omnidirectional video super-resolution. Despite potentially being an obvious extension of conventional VSR, the number of works done in the omnidirectional video super-resolution literature is limited~\cite{liu2022video,baniya2023omnidirectional}. 
Dasari {\it et al.}~\cite{dasari2020streaming} propose a micro-model for super-resolution in {360\textdegree} videos to address bandwidth-related requirements for adaptive video streaming. Their approach focused on enhancing the spatial quality of compressed tiles by passing them through multiple convolution layers and final deconvolution and upsampling. Liu {\it et al.}\cite{liu2020single} introduced a dual network VSR model, named Single and Multi-Frame Recurrent Network (SMFN), for {360\textdegree} videos. One pipeline handles SISR, and the other handles MISR. SMFN has shown better results than conventional VSR models such as EDVR\cite{EDVR} and RBPN~\cite{RBPN}, with targeted training on a {360\textdegree} video dataset. More recently, Agrahari Baniya {\it et al.}~\cite{baniya2023omnidirectional} proposed a spherical signal super-resolution with a proportioned optimisation (S3PO) using recurrent modelling and alignment-free {360\textdegree} feature extraction surpassing state-of-the-art VSR models like BasicVSR~\cite{BasicVSR} in super-resolving omnidirectional videos. Although it has been shown that direct adoption of conventional 2D alignment~\cite{bhandari2021revisiting} and other VSR components into {360\textdegree} VSR models is not the answer to super resolving omnidirectional videos, more work needs to be done to unpack the applicability of 2D deep learning technologies for {360\textdegree} VSR problem.

\subsubsection{3D Video Super-resolution}

3D video super-resolution aims to enhance the resolution of videos captured with depth information and/or multi-view cameras. By leveraging the additional depth, or multi-view information, 3D video super-resolution techniques generate high-resolution videos with improved spatiotemporal details as well as better depth accuracy~\cite{4587703}. Traditional 2D video super-resolution methods are limited in their ability to handle depth information or multi-view data. 3D video super-resolution methods consider both the spatiotemporal and depth dimensions of the video to perform super-resolution~\cite{xie20163d}. These methods need to exploit the geometric correlations between different views or depth maps in a low-resolution video to generate a high-resolution counterpart. Different approaches have been explored in 3D video super-resolution. These include depth-based methods that utilise depth maps to guide the super-resolution process~\cite{li2019high}, view synthesis-based methods that synthesise high-resolution views from the available low-resolution views, and fusion-based methods that combine information from multiple views to generate a high-resolution video~\cite{joachimiak2014view}. With added challenges of occlusions, large disparities between views and depth maps, and limited comprehensive benchmark datasets to be used for model training and development, 3D video super-resolution is, therefore, an emerging field of study.

Although it is a relatively mature field of research, 2D video super-resolution remains an active research area. Numerous unresolved challenges in sequential modelling, along with the emergence of new requirements/applications, have meant that VSR is a progressive and continually evolving research topic. However, often the research conducted in this area lacks the consideration of sequential modelling required for videos and the applicability of these models in terms of the data required as input, model sizes and inference efficiencies. On top of this, as with many other domains implementing deep learning techniques, the composite impact and ability of deep learning methodologies in relation to objective and subjective outcomes remain unexplained. The following sections of this paper aim to investigate the key components and aspects related to the VSR technologies being used, aimed to guide informed model development with explainable methodological choices. 

\vspace{-0.8em}
\section{Overview of Deep Learning-based VSR}
\label{sec:overview}
VSR models employing deep learning predominantly make use of five fundamental methodological components, namely - input, alignment, fusion, refinement and upsampling. Each of these components has a diverse range of methodological options with specific purposes and implications. Fig.~\ref{sec:overview} provides a taxonomic categorisation of the most commonly used options under each component across VSR stages. This taxonomy forms the baseline for the study of literature and methodological analysis conducted in this paper. The technical details, purpose and corresponding substances of each component are discussed in detail in the following sections.
\begin{figure*}[ht]
 \vspace{-3em}

 \centering
 \includegraphics[width = 0.74\textwidth]{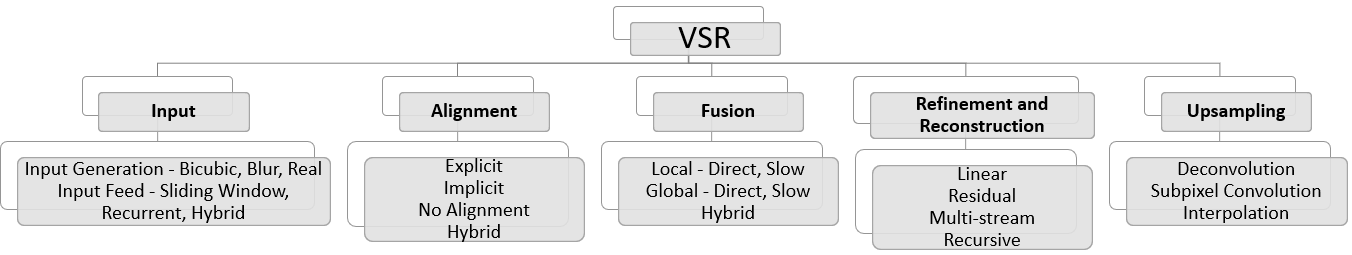}
\vspace{-0.5em}

 \caption{Taxonomy for components across various stages in a VSR model based on the detailed discussion in Sec.~\ref{sec:overview}.}
 \label{Fig:component_taxonomy}
 \vspace{-1em}
\end{figure*}
\vspace{-1em}
    \subsection{Inputs}  
    One of the most important aspects of VSR is the input, as the quality and content of the input directly affect the learning ability of data-driven VSR models and the corresponding outputs generated. In this section, we discuss techniques used for obtaining the input in VSR, including synthetic downgrading and downsampling and real-world LR inputs and how these are fed to VSR models using either a sliding window,  in a recurrent fashion, or via a combination of both.
   
    \subsubsection{Input Generation}
    Synthetic downgrading is a common technique to create LR frames from HR frames for training and evaluating VSR models. This is done by applying various types of quality degradation to the HR frames, such as downsampling, blurring, and the introduction of noise.
    \begin{itemize}
        \item Bicubic downsampling is a simple and widely used technique for creating LR frames from HR frames. Bicubic downsampling involves downsampling the HR frame by a factor of $d$ and applying a cubic interpolation filter across the frame. The filter is typically a square kernel of length $4d+1$ where $d$ is the downsampling factor. Downsampling is obtained by taking a weighted average of every $d$ pixels to obtain $\times d$ reduced spatial resolution. Bicubic downsampling leads to the loss of high-frequency information, reduced spatial pixels and artifacts, such as ringing near edges, namely unwanted oscillations or ripples around sharp edges, resulting in a distorted appearance. The degraded LR image is then used to train deep VSR models in conjunction with its HR ground truth in a supervised manner. 
        \item Blur or Gaussian downsampling is another technique used to create LR frames from HR frames. \rev{Blur or Gaussian downsampling involves convolving the HR frame with a Gaussian kernel ($G_{\sigma}$) with standard deviation \( \sigma \), which effectively blurs the frame and reduces its granular details.} The resulting frame is then downsampled by sampling one pixel out of every $d$ \rev{pixel} to obtain $\times d$ reduced spatial resolution. Compared to bicubic downsampling, aliased artifacts are relatively fewer, and more high-frequency information is preserved when downsampling with a Gaussian filter. This means that restoring high-frequency details from bicubically downsampled low-resolution video is a more challenging task to learn for deep VSR models compared to that with blur downsampling. Moreover, bicubic downsampling is computationally efficient and requires less processing time compared to Gaussian/blur downsampling since blur downsampling involves applying a \rev{computationally intensive convolution operation}.
        \item Real-world LR inputs refer to frames that have been degraded in the real world or closely imitated, including frames captured by low-quality cameras, frames that have been lossy compressed, and frames that have been resized or down-sampled randomly with unknown parameters. A few popular approaches are to apply dynamic blur and image compression techniques, such as JPEG or video compression techniques with random codecs and bitrates. These frames have the advantage of being more representative of the types of degraded frames that users encounter in the real world and, thus, contain degradation that is challenging to revert. Additionally, diversity in degradation can make it challenging to develop a single VSR algorithm that works well for all types of real-world LR videos. Often referred to as blind super-resolution~\cite{pan2021deep, lee2021dynavsr, faramarzi2016blind} or real super-resolution~\cite{chan2022investigating}, the research area aiming to super-resolve real-world low-resolution videos to a high-resolution counterpart remains in its infancy. 
    \end{itemize}

    \subsubsection{Input Feed}
    \begin{itemize}
        \item Sliding-window Feed --- The sliding window approach is a simple and widely used method for inputting frames into a VSR model. It involves dividing a video into overlapping segments of a fixed size and then feeding each segment, also called a ``window'' into the VSR model. The window size is chosen to be small enough to capture temporal variations in the video while also being large enough to provide enough temporal context for the VSR model to learn spatiotemporal correlations. One advantage of the sliding window approach is that it allows the VSR algorithm to use information from past and future frames to generate the output. This can be especially beneficial in cases with significant motion or luminance changes between consecutive frames since the VSR model can establish the spatiotemporal context from other neighbouring frames to better estimate the high-resolution output. The overlapping segments allow flexibility in the input size. While the optimal number of frames per sliding window is subjective to the model and video dataset, limited works have attempted to investigate the effect of the number of frames used in the sliding window input. \rev{Increasing the number of frames from three to five in a sliding window results in improved super-resolution performance; however, increasing the window size beyond five results in consequential computational overhead that eclipses any further result improvement~\cite{RBPN}.}  
        
       \rev{The sliding window approach is computationally expensive as the VSR model must process each frame multiple times due to the overlapping windows.} The overlapping segments may also introduce and propagate repetitive noise into the input across several timestamps, which can negatively impact the quality of the output across the frame series. The number of frames required per target frame resolution has decreased in recent models, with improved temporal information propagation abilities mitigating the need to process videos as disparate pockets of correlated information. However, the temporal sliding window remains the most commonly used input feed mechanism in VSR literature, as observed in Table~\ref{tab:summary}. Despite the popularity of this approach, little is done to treat the temporally dispersed frames in a sliding window differently in relation to the target frame being super-resolved~\cite{EDVR}. Attempts have been made to further segment the sliding window into smaller segments of temporally co-located frames, in either temporal direction with reference to the target frame, and apply different dilations based on temporal distance~\cite{TGA}; however, this approach still assumes a fixed correlation based on temporal distance instead of the actual spatiotemporal dynamics present within each video. Agrahari Baniya \textit{et. al}~\cite{agrahari2023spatiotemporal} have recently demonstrated an alternative sliding-window mechanism with selection measures integrated to perform suitability selection in the input space. Pixel-based and feature-based selections, in this case, have proven to improve the VSR performance showing a direct impact and encouraging consideration of spatiotemporal dynamics in a sliding window for VSR. 

        \item Recurrent Feed --- The recurrent window approach is a more recent method for inputting frames into a VSR algorithm. Recurrent Feed is used with RNNs to feed frames to the VSR model in a recurring manner in a way that is meaningful when used in conjunction with global memory propagation. Depending on the nature of the RNN model, the recurring window can have either unidirectional or bidirectional feeds. The Recurrent feed method usually incorporates the target frame to be super-resolved in confluence with the global memory propagated from the distant timestamps. A simple change in input feed with the use of recurrent feedback has shown to be effective and produce competitive results, often superior~\cite{RRN} to using sliding window frames. This embodies the importance of modelling the task of VSR as a sequential modelling task with recurrent input feed.

        \item Hybrid Feed --- This approach combines elements of both the sliding window and recurrent approaches. Hybrid Feed involves dividing the video into overlapping segments in a similar way to the sliding window approach, but instead of processing each segment independently, the VSR algorithm maintains a recurrent memory state that captures the temporal context across segments. By incorporating both spatial and temporal information, the hybrid approach aims to leverage the benefits of both sliding window and recurrent methods. The hybrid feed allows the VSR algorithm to utilise local spatiotemporal context from previous and future proximal frames within each window while also considering the temporal dependencies propagated across the global timestamps~\cite{BasicVSR}. This has led to improved performance in capturing both short-term and long-term spatiotemporal patterns and producing high-quality super-resolved videos. Ideally, the number of frames per input in a hybrid feed is usually an additional past and/or future frame in reference to the target frame along with the globally propagated recurrent memory~\cite{RSDN, baniya2023online}. Unlike the sliding window method, the need for a large temporal radius for sliding windows is mitigated by global memory propagation. However, the usage of recurrent and hybrid feeds is only applicable to recurrent VSR networks that can maintain recurrent feedback propagation.
        
    \end{itemize}
   \begin{table*}
   \vspace{-3em}
\centering
\caption{Summary of VSR Models and the components used within the respective model based on the taxonomy in Fig.~\ref{Fig:component_taxonomy}.}
\label{tab:summary}
\begin{tabular}{|c|c|c|c|c|c|c|}
\hline
Model & Year & Input Generation & Input Feed & Alignment & Fusion & Refinement \\
\hline
VSRnet~\cite{7444187}& 2016 & Bicubic & Sliding Window & Explicit (MEMC) & Local (Direct) & Linear \\
VESPCN~\cite{caballero2017real}& 2017 & Bicubic & Sliding Window & Explicit (MEMC) & Local (Direct) & Linear \\
SPMC~\cite{SPMCS} & 2017 & Bicubic & Sliding Window & Explicit (MEMC) &Local (Direct) & Residual \\
BRCN~\cite{7919264} & 2018 & Blur, Bicubic & Recurrent & No Alignment & Global (Direct)& Linear \\
FRVSR~\cite{Sajjadi_2018_CVPR} & 2018 & Blur & Recurrent & Hybrid & Global (Direct) &  Residual \\
DUF~\cite{DUF}& 2018& Bicubic&Sliding Window& No Alignment & Local(Direct) & Multistream, Residual\\
FSTRN~\cite{Li_2019_CVPR} & 2019 & Bicubic & Sliding Window & No Alignment & Local (Direct) & Residual  \\ 
3DSRnet~\cite{3dsrnet} & 2019 & Bicubic & Sliding Window & No Alignment & Local (Direct) & Linear \\
TecoGAN~\cite{} & 2019 & Bicubic & Recurrent & Hybrid  & Global (Direct) & Recursive \\
RBPN~\cite{RBPN} & 2019 & Bicubic & Sliding Window & Explicit (MEMC) & Local (Direct) & Residual \\
EDVR~\cite{EDVR} & 2019 & Bicubic & Sliding Window & Implicit (Deformable) & Local (Direct) & Residual \\
RLSP~\cite{RLSP} & 2019 & Blur & Recurrent & No Alignment & Global (Direct) & Linear \\
TDAN~\cite{TDAN} & 2020 & Blur, Bicubic & Sliding Window & Implicit (Deformable) & Local (Direct) & Linear \\
TGA~\cite{TGA} & 2020 & Blur & Sliding Window & Explicit(MEMC) & Local (Slow) & Linear \\
RSDN~\cite{RSDN}& 2020 & Bicubic & Recurrent & No Alignment & Global (Direct) &  Residual, Multistream \\
RRN~\cite{RRN}& 2020 & Bicubic & Recurrent & No Alignment & Global (Direct) &  Residual\\
MuCAN~\cite{10.1007/978-3-030-58607-2_20}& 2020 & & Sliding Window & No Alignment & Local (Slow) & Linear \\
D3D~\cite{9153920}& 2020 & Bicubic &Sliding Window & Implicit (Deformable) & Local (Direct) & Residual\\
MSFFN~\cite{9351768} & 2021 & Bicubic & Recurrent & Implicit (Deformable) & Global (Slow) & Residual \\
STMN~\cite{ZHU2021107619} & 2021 & Bicubic & Sliding Window & No Alignment & Global (Direct) & Residual \\
EVSRNet~\cite{Liu_2021_CVPR} & 2021 & Bicubic & Sliding Window & No Alignment & Local (Direct) & Residual \\
FDAN~\cite{lin2021fdan} & 2021 & Blur & Sliding Window & Implicit (Deformable)  & Local (Direct) & Residual \\
 &  &  &  &  Explicit (Flow) &  &  \\
 BasicVSR~\cite{BasicVSR} & 2021 & Blur, Bicubic & Recurrent & No Alignment & Global (Direct) &  Residual \\
 IconVSR~\cite{BasicVSR} & 2021 & Blur, Bicubic & Hybrid & Hybrid & Hybrid (Direct) &  Residual \\
 GOVSR~\cite{yi2021omniscient} & 2021 & Blur & Hybrid & No & Hybrid (Direct) & Residual\\
MSHPFNL~\cite{9279273} & 2022 & Blur, Bicubic & Sliding Window & No Alignment & Local (Progressive) & Residual \\
TTVSR~\cite{Liu_2022_CVPR} & 2022 & Blur, Bicubic & All Frames & Explicit (MEMC) & Global (Direct) & Residual \\
BasicVSR++~\cite{chan2022basicvsr++} & 2022 & Blur, Bicubic & Recurrent & Hybrid & Global (Direct) & Residual \\
PSRT~\cite{NEURIPS2022_ea4d65c5} & 2022 & Bicubic & Sliding Window & Explicit (MEMC), & Local (Direct) & Residual \\ 
&  &  &  &  Implicit (Deformable) &  &  \\ 
R2D2~\cite{baniya2023online} & 2023 & Blur & Hybrid & Hybrid & Hybrid (Direct) & Residual, Multistream \\
R2D2-{\it lite}~\cite{baniya2023online} & 2023 & Blur & Hybrid & No & Hybrid (Direct) & Residual, Multistream \\
\hline
\end{tabular}
\vspace{-1.5em}
\end{table*}

\vspace{-1.5em}   
    \subsection{Alignment}
    Alignment refers to the process of aligning low-resolution neighbouring frames or features with their corresponding target counterparts being super-resolved. There are several types of frame/feature alignment methods used in VSR, including:
    \subsubsection{Explicit Alignment}
    Motion Estimation and Motion Compensation (MEMC) based alignment is a widely used method for aligning frames and features explicitly in VSR. The method uses dense optical flow estimated between neighbouring and target frames/features to align them. It computes the motion vector for each pixel by analysing the changes between consecutive frames, assuming that the intensity pattern observed in a frame remains constant over time and any variation is due to the movement of objects or the camera. The flows are then used to warp the neighbouring frames/features to match the position of the target counterpart. This method is efficient and has been widely used~\cite{caballero2017real}. However, it has some limitations when dealing with complex, fast, and large motions or significant changes in luminance. In these cases, the estimated optical flow may not be accurate, leading to misalignment and degraded VSR performance. To mitigate some of the limitations, traditional methods of flow estimations are being increasingly replaced with learning-based motion estimation methods~\cite{spynet}, such methods proving to be more robust to various motion types.
    
    \subsubsection{Implicit Alignment}
    Deformable convolution~\cite{deformConv} is a method that allows convolutional kernels to adapt to the shape of the input feature map. This method has been applied to align frames and features implicitly in VSR and has shown promising results~\cite{chan2021understanding}. Deformable convolution learns the geometric variations and motion of objects in the video and deforms the kernel to adjust it accordingly. The method thus provides implicit alignment capabilities. Deformable convolution is robust to changes in motion and luminance; however, it requires a large number of parameters and thus increases the computational overhead of the VSR model.
    
    \subsubsection{No Alignment} Instead of aligning frames, features are extracted directly from frames using either 2D, 3D or recurrent neural networks without any alignment. These methods have the advantage of being computationally efficient and do not require any explicit or implicit alignment overhead~\cite{Sajjadi_2018_CVPR}. However, no alignment as a strategy may not be as effective as alignment-based methods in handling complex motions since the VSR model solely relies on the effectiveness of spatial or spatiotemporal features extracted by the 2D or 3D/recurrent convolution layers, respectively. Without the alignment to guide temporal coherence in model outputs, non-alignment methods are prone to inconsistencies, with the notable exception of RNNs, which maintain global memory propagations aimed at assisting temporal modelling. However, unwanted noises or changes in the recurrent feedback due to sudden large motion or luminance change are likely to propagate across a longer temporal radius if not interrupted and rectified. \rev{In addition to the use of 2D, 3D and recurrent networks for no alignment, non-local operations using Gaussian functions and dot products, pixel operations like RGB difference maps and forward and backward feature differences are used for finding relationships and correlations between frame pairs or feature pairs~\cite{PFNL,9279273,xiao2023local}}. However, this approach introduces additional computational costs and is more often than not outperformed by recurrent methods. 
 
    \subsubsection{Hybrid Alignment} More recently~\cite{yi2021omniscient,BasicVSR}, to mitigate the shortcomings of the no-alignment approach while leveraging its efficiency, hybrid methods combine alignment and non-alignment techniques to handle different types of motion and improve the accuracy of alignment. These methods provide VSR models with the flexibility to leverage the robustness of implicit/explicit alignment methods in the event of large or sudden motion changes while benefiting from the efficiency of non-alignment components when motion or luminance changes are less significant. ~\rev{An example is the patch alignment technique in transformers which uses image alignment, feature alignment and deformable convolution to align image patches instead of pixels to mitigate the computational overhead of pixel alignment and prevent the impact of alignment on attention \revb{capabilities}~\cite{NEURIPS2022_ea4d65c5}. Similarly, a memory-augmented attention module has been used to maintain a global memory of the entire training set learned as parameters of the network while using regular non-local attention to query current frame features in the global memory bank to memorise and utilise general video details during the super-resolution training~\cite{yu2022memory}.} By combining the benefits of different alignment methods, hybrid approaches aim to enhance the overall performance of VSR models. This has only been introduced in recent years with only reported usage in recurrent models using hybrid input feeds discussed earlier~\cite{baniya2023online}.  
    
    \vspace{-1em}
    \subsection{Fusion}
    Several types of feature/frame fusion are used in VSR models for integration or combination of multiple low-resolution frames or features before refining, reconstruction, and upsampling. Some of the most common methods include: 
    \subsubsection{Local Information Fusion}
    Local fusion focuses on utilising the information from the immediate surrounding pixels or frames/features in a local neighbourhood to enhance the resolution of a specific area or a specific frame. These methods typically involve the use of convolution layers to extract features from the low-resolution (LR) frame(s)/feature(s) and use the subsequent features for further processes.
    \begin{itemize}
        \item Direct Fusion: Direct fusion methods involve concatenating the target LR input frame/feature with the features extracted from a local neighbourhood and then passing the concatenated features through a series of layers for further refinement, reconstruction and upsampling. In order to fuse the information, either concatenation along the depth (most common) or addition along the co-located spatial dimension is performed. Direct fusion results in the abstraction of the temporal variations and dilutes the correlation differences that might be preset in the information. The expectation is for deep model layers to be able to extract meaningful features from the fused information. Despite its simplicity, this approach has been most commonly used, as shown in Table~\ref{tab:summary}, and has proven to be effective with the ability of deep learning to automatically extract features of interest in relation to the super-resolution task.

        \item Slow Fusion: Slow fusion methods involve using multiple stages of feature extraction and fusion in series, where each convolution layer extracts features from the previous layer and fuses it with other relevant information. This approach allows for a more gradual and fine-grained fusion of the LR input with the extracted features allowing for a better hierarchical feature representation, which can result in improved HR output. Although fusions from 3D models and elongated progressive fusions are categorised differently to slow fusion in some of the literature~\cite{PFNL,9279273}, given the nature of fusion irrespective of the number of steps, or type of convolution involved, we categorise multi-stage fusion methods under slow fusion for simplicity's sake. The slow fusion methods also make use of either concatenation or aggregation operations to fuse the information (frame(s) and/or feature(s)) gradually. Unlike direct fusion, this approach aims to overcome the dilution and abstraction of spatiotemporal variations in the information and establish gradual relationships between different sets of information. This is particularly beneficial for methods using sliding window feed since it helps establish a hierarchical relationship between the temporally dispersed frames within a sliding window. The direct cost of the additional computation in relation to the magnitude of the improved result is, however, not fully studied in the literature. 
    \end{itemize}
    \vspace{-0.5em}
    \subsubsection{Global Information Fusion}
    Global information fusion focuses on utilising information from a larger temporal dimension to enhance the resolution of a specific frame. This class of methods typically involve the use of recurrent feedback to extract features from a wider video frame sequence and then use these features to generate the HR output. This fusion technique also makes use of either concatenation or aggregation to fuse the local frame(s) and/or feature(s) with global memory. The process of fusion can either be direct or slow, in a similar way to the local fusion methods. However, unlike the local approach, the features extracted will represent the temporal context across the frame sequence allowing for better sequential modelling resulting in temporal coherence. This approach is only used with recurrent models making use of a recurrent input feed with no alignment discussed earlier.
    
    \subsubsection{Hybrid Information Fusion}
    Hybrid information fusion methods combine both local and global information to take advantage of the strengths of both approaches. Different frequency details and spatiotemporal relevance can be presented by combining the two approaches~\cite{baniya2023online}. This allows for a more comprehensive fusion strategy that can capture both local details and global temporal coherence. The global and local information, in this case, is fused either directly or slowly using concatenation or aggregation.  The most common approach is to use hybrid fusion with earlier discussed hybrid input feed and the hybrid alignment techniques. For example, two sets of information obtained from hybrid alignment (explicit alignment in a sliding window and no alignment in recurring feed) can be fused directly (or slowly) over multiple layers. 

   \vspace{-1em}
    \subsection{Refinement and Reconstruction}
    \subsubsection{Linear}
    Linear refinement is a process that entails utilising a solitary linear pipeline consisting of convolution layers to extract valuable features from the amalgamated input features. This method of refinement and reconstruction is simple but also limited in effectiveness. Each future layer depends on features propagated from previous layers, unable to mitigate any noisy or irrelevant features from being propagated, leading to a direct impact on the generated output. Moreover, increasing the depth of the neural models leads to the fading of information propagated during backpropagation which hinders the learning ability of these models. Linear refinement models are increasingly uncommon as the advancements in deep learning space continue to be accelerated.
    
    \subsubsection{Residual}
    Residual refinement involves using residual blocks with skip connections for propagation. Residual blocks are designed to allow the network to propagate extracted features in conjunction with input to the respective layers to mitigate the shortcomings of its linear counterparts. This approach specifically helps mitigate the propagation of noise and unwanted information induced at any layer while preventing backpropagation info from fading and reaching deeper layers. The residual approach is the most common refinement approach currently in VSR literature for the reconstruction of fused features, as is evident in Table~\ref{tab:summary}.

    \subsubsection{Multi-stream}
    Multi-stream refinement uses multiple pipelines to propagate and refine features. Each pipeline can learn differently to extract different features from the input, and the output from each pipeline is combined to generate the final output. The individual pipelines within the multi-stream refinement approach can be linear or residual. This approach allows the network to learn more complex feature representation with dedicated pipelines; however, it can significantly increase the model complexity and size. This approach is less common in the literature, but it has been shown to be effective when used strategically. Recent evident usage of this approach was observed in RSDN~\cite{RSDN} and R2D2~\cite{baniya2023online}, where the two pipelines were used to propagate structure and detail information and local and global information, respectively. The effectiveness of this approach over a single pipeline of residual blocks was highlighted in both works.

    \subsubsection{Recursive}
    Recursive refinement is used to refine the fused features in a recursive manner by iteratively applying refinement steps. This approach involves repeatedly passing the fused features through refinement modules to improve the quality of the final feature gradually. Each refinement module can be designed using linear or residual recurrent architectures. Recursive refinement allows for iterative fine-tuning of the features, enhancing their representation and reducing artifacts. However, this approach can be computationally expensive due to the repeated refinement steps and may therefore require careful tuning to avoid over-fitting or convergence issues. Recursive refinement is very uncommon because of its repetitive nature and is often replaced with linear or residual counterparts used in conjunction with a recurrent neural network instead of recursively refining the features for every timestamp.

       \vspace{-1em}
    \subsection{Upsampling}
    Upsampling refers to the process of increasing the spatial resolution of low-resolution feature(s) or frames to obtain high-resolution counterparts. Various methods are used for upsampling, including the following:
    
    \subsubsection{Transposed Convolution (Deconvolution)} Transposed convolution, also known as deconvolution, is a popular method for upsampling in deep learning that uses learned kernels to upsample the feature maps by convolving them with the input~\cite{lai2017deep,DUF}. The learnable kernels are similar to convolutional kernels but with the dimensions reversed. During the deconvolution process, the kernel slides over the input feature map, and instead of performing a dot product as in standard convolution, it computes the outer product of the kernel with the input. This process effectively spreads the activations and increases the resolution of the feature map. Deconvolution layers often include parameters such as stride and padding, similar to convolutional layers. The stride determines the step size of the kernel during sliding, and the padding controls the size of the output feature map. By adjusting these parameters, the resolution of the output feature map can be controlled. However, deconvolution has a tendency to generate checkerboard-like artifacts in the output. It refers to a visual artifact that can occur when performing upsampling or transposed convolution operations which are characterised by an exhibition of a pattern resembling a checkerboard, with alternating square regions of high and low intensity, negatively impacting the visual quality of the super-resolved video. To mitigate this issue, various regularisation techniques, such as adjusting the stride or dilation rate, or using skip connections, can be employed. Stride refers to the step size or the distance by which a filter/kernel moves across an input image during convolution or pooling operations. It determines the amount of overlap or spacing between successive applications of the filter. While dilation, also known as atrous convolution, is a technique used to increase the receptive field of a convolutional layer without increasing the number of parameters. The receptive field refers to the region of the input that a neuron in the convolutional layer "sees" when performing convolution. In a regular convolution operation, each filter/kernel is applied to a local region of the input data, and the size of the output feature map is determined by the size of the filter and the stride used. The receptive field of a neuron in the convolutional layer depends on the filter size and the number of layers it has passed through in the network. With dilation, gaps or spaces are introduced between the kernel elements, effectively increasing the stride of the convolution. This results in an expanded receptive field without altering the size of the filter. The dilation rate controls the spacing between the elements of the kernel. Similarly, skip connection refers to a mechanism that allows the direct flow of information from one layer to another, bypassing intermediate layers in a deep neural network architecture.

     \subsubsection{ Pixel Shuffle (Subpixel Convolution)} Pixel shuffle, or subpixel convolution, is a technique for upscaling or increasing the spatial resolution that rearranges the elements of low-resolution feature maps to form high-resolution feature maps by performing depth-to-space transformations~\cite{ESPCN}. The rearrangement takes place in the channel dimension (depth) of the feature map. For example, a feature map with dimensions [batch size, channels, height, width], where ``channels'' denotes the number of feature channels, and ``height'' and ``width'' represent the spatial dimensions, is reshaped so that each channel is divided into smaller subgroups of $d^2$ elements where $d$ is the upsampling factor. These subgroups are spatially rearranged to form a higher-resolution feature map. Specifically, the $d^2$ elements from each subgroup are combined together to form individual pixels in the output feature map, which has a higher resolution. This method does not introduce any additional learnable parameters, which in turn contributes to its computational efficiency and popularity in super-resolution models with effective results.
        
       \subsubsection{ Bilinear or Bicubic Interpolation} Bilinear or bicubic interpolation methods are simple and computationally efficient upsampling techniques. These methods estimate the values of new pixels based on the surrounding pixels in the low-resolution frames. The filter kernels utilised in traditional bilinear or bicubic upsampling are static and unchangeable, with the only adjustment being the kernel's position based on the location of the newly generated pixel in the upsampled frame. When performing $\times 4$ upsampling, a fixed set of 16 kernels is employed in these conventional methods. While the kernels are fast, they rarely fully restore sharpness and preserve fine textures in the resulting frame regions. Bilinear or bicubic interpolation methods are often used in conjunction with residual features from the deep VSR models. Instead of relying on the deep model to create every pixel, the residual approach allows the model to focus on the details that can be added on top of either the bilinear or bicubic interpolated low-resolution target frame. This method remains the most common approach to upsampling. However, the upsampling of the residual feature, in this case, is usually done with the pixel shuffle operation.

\section{Architecture, Training and Datasets }
\label{sec:deep_tech}
    \subsection{Network Architecture}
    The deep network architecture plays a crucial role in enhancing the learning capability and determining the components employed in VSR models. It also governs the overall modelling approach for video super-resolution. Depending on the chosen architecture, VSR models can learn to super-resolve videos either sequentially or non-sequentially. In this section, we delve into the prevalent deep learning architectures frequently utilised in the development of VSR models, as outlined in the provided Table~\ref{tab:deep_res}.

        \subsubsection{Non-sequential}
            \begin{itemize}
                \item {\it 2-Dimensional Convolutional Neural Networks (2D CNNs)} have been the cornerstone of image-related tasks due to their ability to learn hierarchical features from spatial data. In the context of VSR, 2D CNNs treat each frame independently, essentially performing image super-resolution on every frame by making use of LR frames available in a sliding window. While computationally efficient, this approach neglects the temporal correlations between consecutive frames, which are crucial for video. The convolution filters in 2D CNNs extract spatial features from each frame or fusion of frames and apply activation functions such as ReLU, sigmoid or tanh to introduce non-linearity. These features are then refined and upsampled to generate super-resolved outputs. 2D CNN-based VSR models struggle to handle videos with substantial motion or dynamic scenes due to a lack of temporal information and have become uncommon.

\begin{table*}
\vspace{-4.6em}
\centering
\caption{\textbf{Comprehensive summary of key learning aspects of VSR models in the literature} and their reported objective performance on different test datasets. Y and RGB indicate the channel of HR output used for computing quality. In the case of models with different input degradation, as reported in Table~\ref{tab:summary}, their corresponding bicubic degradation-related results are reported. \rev{Size represents model parameters in millions}. \revb{The application represents the applicability of the model based on input feed.}}
\scriptsize
\label{tab:deep_res}
\begin{tabular}{|c|c|c|c|c|c|c|c|c|}
\hline
{\bf Model}  & {\bf Network Arch.} & {\bf Loss Function} & {\bf Training Dataset} & {\bf Test Dataset} & {\bf PSNR/SSIM} & \rev{{\bf Size}} & \rev{{\bf Application}}\\
\hline
VSRnet~\cite{7444187}& 2D CNN& Euclidian Distance & Myanmar & Vid4 (Y) & 24.84 / 0.7049 &\rev{0.27}&\rev{Online}  \\ \hline
VESPCN~\cite{caballero2017real}&  2D CNN & Weighted (L2, Huber) & CDVL & Vid4 (Y) & 25.35/0.7557 &\rev{0.88}&\rev{Online}\\ \hline
DRVSR~\cite{SPMCS} &  Encoder-Decoder & Weighted (Warp- & SPMCS & SPMCS (Y) & 29.89 / 0.84&\rev{2.17}&\rev{Online} \\
 &  & ing + Euclidean) & &  & & & \\\hline
BRCN~\cite{7919264} &  3D-Bidirec. RNN& L2 loss & YUV25 & Vid4 (RGB) & 24.43/0.6334&\rev{-}&\rev{Offline} \\ \hline
FRVSR~\cite{Sajjadi_2018_CVPR} &  RNN & Weighted (MSE, & Custom (vimeo.com) & Vid4 (Y) & 26.69/0.8220&\rev{2.81}&\rev{Online} \\
  &  &  Warping Error)&  &  & & & \\\hline
DUF~\cite{DUF} &3D CNN & Huber loss & Custom  & Vid4 (Y) & 27.34 / 0.8327 &\rev{5.82}&\rev{Online}\\ \hline
FSTRN~\cite{Li_2019_CVPR}  & 3D CNN & Charbonnier loss & YUV25 & -&-&\rev{-}&\rev{Online} \\\hline
3DSRnet~\cite{3dsrnet}  & 3D CNN & L2 Loss & Custom - smallSet & Vid4 (RGB) & 25.46/0.7498&\rev{0.11}&\rev{Online} \\
&&&Custom - largeSet &Vid4 (RGB)& 25.71/0.7588&\rev{}&\\\hline
RBPN~\cite{RBPN} &  Encoder-Decoder & L1 loss & Vimeo-90k & Vid4 (Y) & 27.12/0.818&\rev{12.2}&\rev{Online} \\
&&&&SPMCS-11 (Y)& 30.10/0.874&\rev{}&\\
&&&&Vimeo-90k (Y)& 37.16/0.9566&\rev{}&\\\hline
EDVR~\cite{EDVR} &  2D CNN & Charbonnier loss & REDS & REDS4(Y) & 31.09/0.8800&\rev{20.60}&\rev{Online} \\
&&&Vimeo-90k&Vid4 (Y)& 27.35/0.8264&\rev{}&\\
&&&Vimeo-90k&Vid4 (RGB)& 25.83/0.8077&\rev{}&\\\hline
RLSP~\cite{RLSP} &  RNN & MSE & Vimeo-90k & Vid4 (Y) & 27.55/0.838&\rev{4.21}&\rev{Online} \\ \hline
TDAN~\cite{TDAN} &  2D CNN & Weighted (Alignment MSE & Vimeo-90k & Vid4(RGB) & 26.42/0.789&\rev{1.97}&\rev{Online}\\
&  &+ Supre-resolution MSE) &   & SPMCS-30 (RGB)& 30.38/0.854&\rev{}&\\\hline
TGA~\cite{TGA} &  2D and 3D CNN & L1 loss & Vimeo-90k & Vid4 (Y) & 27.59/0.8419&\rev{5.8}&\rev{Online}\\
&&&&Vid4 (RGB)& 26.10/0.8254&\rev{}&\\\hline
RSDN~\cite{RSDN} &  RNN & Weighted (Charbonnier  & Vimeo-90k & Vid4 (Y) & 27.92/0.8505&\rev{6.19}&\rev{Online} \\
&&&+ L2 )&Vid4 (RGB)& 26.43/0.8349&\rev{}&\\\hline
RRN~\cite{RRN}&  RNN & L1 loss &  Vimeo-90k & Vid4 (RGB)& 26.16/0.8209&\rev{3.4}&\rev{Online}\\
& &  &   & SPMCS (RGB)& 28.28/0.8690&\rev{}&\\
& & &    & UDM10 (RGB)& 37.03/0.9534&\rev{} &\\\hline
MuCAN~\cite{10.1007/978-3-030-58607-2_20} &  Encoder-Decoder & Weighted (Charbonnier & REDS & REDS & 30.88/0.8750&\rev{-}&\rev{Online}\\
&  & + Edge Aware Loss)&  Vimeo-90k & Vimeo-90k (RGB) & 35.49/0.9344 &\rev{}&\\\hline
D3Dnet~\cite{9153920} &  3D CNN & L2 loss & Vimeo-90k & Vid4 (RGB) & 26.52/0.799&\rev{2.58}&\rev{Online} \\ 
& & & &   Vimeo-90k (RGB) & 35.65/0.9330&\rev{}& \\
& & & &   SPMC-11 (RGB) & 28.78/0.8510&\rev{}& \\\hline
MSFFN~\cite{9351768} & RNN & Charbonnier loss & Vimeo-90k & Vid4 (Y) & 27.23/0.8218 &\rev{}&\rev{Online}\\
& & &   & Vimeo-90k (Y) & 37.33/0.9467 &\rev{8.5}&\\
& & &   & SPMC-11 (Y) & 30.13/0.8769 &\rev{}&\\\hline
STMN~\cite{ZHU2021107619} &   3D CNN & MSE & Custom (Vimeo.com  & Vid4(Y) & 25.90/0.7878&\rev{-} &\rev{Online}\\
 &  &    &  + 699pic.com) &  &  &\rev{}&\\\hline
EVSRNet~\cite{Liu_2021_CVPR} &  2D CNN & L1 Loss & REDS & REDS4(Y) & 27.85/- &\rev{-}&\rev{Online}\\ \hline 
FDAN~\cite{lin2021fdan} &  2D CNN & L1 loss & Vimeo-90k & Vimeo-90k-T (Y) & 37.75/0.9522 &\rev{8.97}&\rev{Online}\\ 
& & & &  Vid4 (Y) & 27.88/0.8508&\rev{}& \\
& & & &  UDM10 (Y) & 39.91/0.9686&\rev{}& \\ \hline
BasicVSR~\cite{BasicVSR} &  Bidirectional RNN & Charbonnier loss & REDS & REDS4 (RGB) & 31.42/0.8909&\rev{6.3}&\rev{Offline} \\
& & &  Vimeo-90k & Vimeo-90k-T (Y) & 37.18/0.9450&\rev{}&  \\
& & &  Vimeo-90k & Vid4 (Y) & 27.24/0.8251&\rev{}&  \\ \hline
IconVSR~\cite{BasicVSR} &  Bidirectional RNN & Charbonnier loss & REDS & REDS4 (RGB) & 31.67/0.8948&\rev{8.7}&\rev{Offline} \\
& & &  Vimeo-90k & Vimeo-90k-T (Y) & 37.47/0.9476&\rev{}&  \\
& & &  Vimeo-90k & Vid4 (Y) & 27.39/0.8279 &\rev{}& \\ \hline
GOVSR~\cite{yi2021omniscient} &  Bidirectional RNN & Charbonnier &  MM522 & Vid4 (Y) & 28.41 / 0.8724&\rev{1.90}&\rev{Offline} \\ 
& & &  MM522 & UDM10 (Y) & 40.14/0.971&\rev{}& \\ \hline
MSHPFNL~\cite{9279273}&  GAN & Weighted (adversarial, & MM522 & Vid4 (Y) & 27.70/0.8472&\rev{7.77}&\rev{Online}\\
& &  perceptual,frame variation)& MM522 & UDM10 (Y) & 39.59/0.9676&\rev{}& \\
& & &  Vimeo-90k & Vimeo-90k-T (Y) & 36.75/0.9406&\rev{}& \\\hline
TTVSR~\cite{Liu_2022_CVPR} & Transformer & Charbonnier loss & REDS & REDS4 (Y) & 32.12/0.9021 &\rev{6.8}&\rev{Online}\\
& &  & Vimeo-90k & Vid4 (Y) & 28.40/0.8643&\rev{}&  \\ 
& &  & Vimeo-90k & UDM10 (Y) & 40.41/0.9712 &\rev{}& \\
& &  & Vimeo-90k & Vimeo-90k-T (Y) & 37.92/0.9526&\rev{}&  \\\hline
BasicVSR++ &  Bidirectiona RNN &  Charbonnier loss & REDS & REDS4 (RGB) & 32.39/0.9069&\rev{7.3}&\rev{Offline} \\
~\cite{chan2022basicvsr++}& &  & Vimeo-90k & Vimeo-90k-T (Y) & 37.79/0.9500&\rev{}&   \\
& &  & Vimeo-90k & Vid4 (Y) & 27.79/0.8400&\rev{}&  \\ \hline
PSRT~\cite{NEURIPS2022_ea4d65c5} &  Transformer & Charbonnier loss & REDS & REDS4 (Y) &  31.32/0.8834&\rev{-}&\rev{Online} \\ \hline
R2D2~\cite{baniya2023online} &  RNN &  L1 Loss & Vimeo-90k & Vid4 (Y) & 28.13/0.9244&\rev{8.25}&\rev{Online}  \\
& &  &  & UDM10 (Y) & 39.53/0.9670&\rev{}&  \\
& &  &  & SPMCS11 (Y) & 30.71/0.8920&\rev{}&   \\ \hline
R2D2-{\it lite}~\cite{baniya2023online} & RNN &  L1 Loss & Vimeo-90k & Vid4 (Y) & 28.05/0.8552&\rev{6.85} &\rev{Online}  \\
& &  &  & UDM10 (Y) & 39.38/0.9661&\rev{}&  \\
& &  &  & SPMCS11 (Y) & 29.91/0.8758 &\rev{} & \\ \hline
\end{tabular}
\vspace{-2em}
\end{table*}

                \item {\it 3-Dimensional Convolutional Neural Networks (3D CNNs)}  extend the concept of 2D CNNs to incorporate temporal information by operating on spatiotemporal data cubes. By applying filters across three dimensions: width, height, and time jointly, these models process both spatial and temporal dimensions producing 3D feature maps and effectively capturing motion patterns and temporal dependencies within a given video segment (usually a sliding window)~\cite{3dsrnet,DUF,li2019high,9153920}. This makes 3D CNNs well-suited for tasks involving dynamic scenes and motion-based artifacts. However, the use of 3D convolutions increases the model's computational complexity and memory requirements, making it challenging to scale to longer video sequences or low-resource devices. Moreover, the limited temporal context confined to a fixed temporal radius in a sliding window does not allow these models to exploit the temporal correlation across the video frame series fully.
                \item {\it Generative Adversarial Networks (GANs)} have shown impressive results in various image synthesis tasks, including single-image super-resolution~\cite{wang2020deep}. In the context of VSR, GAN-based models consist of a generator network that produces high-resolution video frames and a discriminator network that distinguishes between real high-resolution frames and generated synthetic ones~\cite{FeatGAN,9279273}.  Adversarial training encourages the generator network to produce visually realistic results. However, training GAN-based VSR models can be challenging due to the inherent instability of GAN training. Training GANs often require extensive hyperparameter tuning and a large quantity of training data to achieve high-quality results. Additionally, GAN-based models may suffer from mode collapse, where \revb{they fail} to capture the full diversity of the training data distribution and instead \revb{generate} a limited set of outputs, often repetitive or unrepresentative of the entire dataset. It occurs when the model ignores or collapses multiple modes (distinct patterns or clusters) in the data and focuses on generating only a few dominant modes~\cite{jabbar2021survey}. GAN-based models' ability to capture temporal information is also limited, similar to that of 2D CNNs, as GAN generators apply 2D convolution operations to extract features from frames in a sliding window, enabling only spatial features capture~\cite{FeatGAN,9279273}.
                \item {\it Encoder-Decoder architectures} have been successful in various image processing tasks, and their application to VSR is likewise promising. These networks consist of an encoder that compresses the low-resolution video frames into a latent representation and a decoder that reconstructs the high-resolution output. Although encoder-decoder networks are known to be used for sequential modelling with RNNs~\cite{cho2014learning}, or their variants such as LSTM~\cite{malhotra2016lstm} or gated recurrent unit (GRU)~\cite{jin2021deep} in other domains, in VSR, encoder-decoder networks are commonly been used with sliding windows of fixed length, employing 2D convolution layers for feature extraction and fusion enabling them to learn spatial information only~\cite{RBPN,10.1007/978-3-030-58607-2_20}. Thus, encoder-decoder networks struggle to fully capture long-range temporal dependencies, limiting their performance in videos with complex temporal patterns.
                \item {\it Transformers} have revolutionised natural language processing tasks and have recently attracted attention in computer vision applications. In the context of VSR, transformers can effectively model temporal relationships between frames in a sliding window by employing self-attention mechanisms~\cite{sun2020attention}. This allows them to weigh the importance of different frames dynamically, capturing temporal variations and correlations within a sliding window. Although transformer use is \revb{limited in} VSR, they have shown promising results in handling complex temporal patterns and achieving state-of-the-art performance in various other video-related tasks such as object detection, recognition, captioning, segmentation and anomaly detection~\cite{han2022survey}. However, the computational complexity of transformers is greater than traditional CNN-based architectures, which may hinder their real-time deployment on resource-constrained devices.
            \end{itemize}
        \subsubsection{Sequential}
            \begin{itemize}
                \item {\it Unidirectional RNNs},  using LSTM, GRU or residual blocks, process video frames sequentially from past to future. They have the ability to model temporal dependencies and retain long-term information through recurrent connections. However, their main limitation is the restricted context they consider since they only have access to past frames during training. As a result, unidirectional RNN-based VSR models may struggle to capture complex temporal patterns in videos where bidirectional interactions between frames are prevalent. Moreover, the amount and relevance of the temporal context available for earlier timestamps in a video sequence are inadequate, often causing subpar performance for the initial frames and a gradual improvement over time~\cite{RRN}. Nevertheless, RNNs have proven to be one of the more effective architectures in modelling the VSR task sequentially. Recent advances over vanilla RNNs tend to address the vanishing gradient issue and information availability for earlier frames by using hybrid input feed and alignments~\cite{baniya2023online,yi2021omniscient}. \rev{Further integrating learning techniques such as LSTM residual blocks~\cite{zhu2019residual,8579237}, latent representation~\cite{RLSP}, feature attention~\cite{baniya2023omnidirectional} and multi-stream refinement~\cite{RSDN,baniya2023online} have resulted in improved RNN performance for VSR, overcoming earlier challenges and performing competitively with significantly reduced computation and number of frames.}

\begin{figure*}[ht]
\vspace{-2em}
 \centering
 \includegraphics[width = \textwidth]{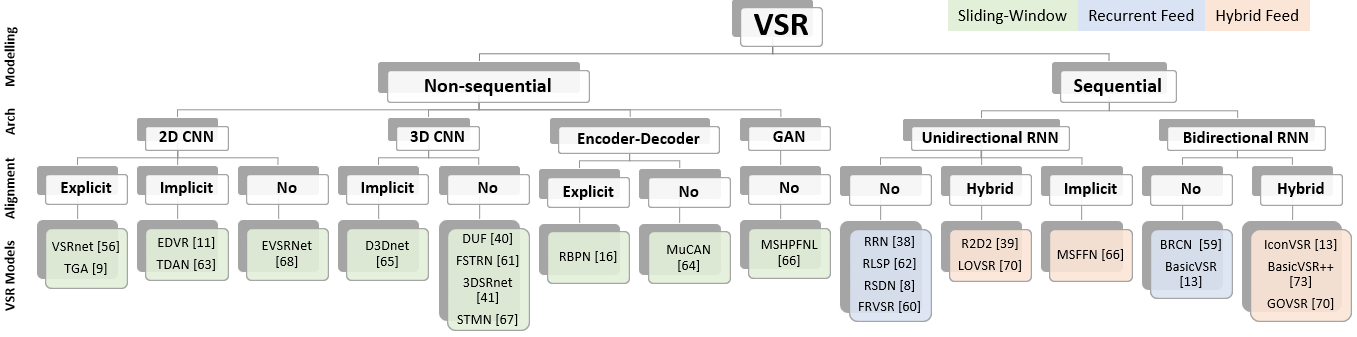}
 \vspace{-2em}
 \rev{\caption{Highlight of VSR models from the literature based on discussions in Sec.~\ref{sec:overview} and ~\ref{sec:deep_tech}.}}
 \label{Fig:models}
 \vspace{-1.5em}
\end{figure*}

      \item {\it Bidirectional RNNs} aim to overcome the limitations of unidirectional RNNs by processing video frames in both forward and backward temporal directions simultaneously. By doing so, Bidirectional RNNs can capture information from past and future frames, providing a more comprehensive context for making super-resolution predictions. Bidirectional RNNs are well-suited for tasks involving bidirectional motion or events where future information significantly influences the current frame's super-resolution. However, bidirectional RNNs introduce higher computational costs compared to unidirectional RNNs and need entire video frames to be available during inference, making this approach unfit for several applications related to real-time and online VSR applications.
            \end{itemize}

     \subsubsection{Network Selection}  
     For computational efficiency and real-time applications, 2D CNNs and Encoder-Decoder networks are preferred over 3D CNNs and Transformers. However, when capturing long-range dependencies and handling complex temporal patterns is crucial to the  VSR task, and where efficiency is a priority, unidirectional RNNs are more suitable choices. GAN-based VSR models may suffer from training instability and therefore require a larger training dataset, while traditional CNN-based and RNN-based architectures offer more stable training with the potential to perform well using smaller training datasets. Additionally, 3D CNNs and Transformers have greater memory requirements, making 2D CNNs and Encoder-Decoder networks preferable for memory-limited scenarios. Applications with no real-time and online constraints can leverage the full potential of temporal modelling from bidirectional RNNs in offline settings.

     The taxonomy presented in Fig.~\ref{Fig:component_taxonomy} in combination with the learning elements discussed in Sec.~\ref{sec:deep_tech} and presented in Table~\ref{tab:deep_res} have been combined and summarised in Fig.~\ref{Fig:models} which provides the merged synopsis of the learning-based modelling technique used by VSR models in literature along with their respective network architecture, alignment technique and input feed mechanism. This helps establish an association between the component methodologies discussed in Sec.~\ref{sec:overview} in relation to the deep learning techniques discussed in Sec.~\ref{sec:deep_tech}. It is evident from Fig.~\ref{Fig:models} that non-sequential modelling remains the most common approach to VSR solution development in the literature. It is also evident that sequential modelling using RNNs is becoming popular with no use of explicit or implicit alignment techniques except for hybrid alignment. No alignment technique also appears common for models deploying spatiotemporal feature extraction capabilities such as 3D CNNs and RNNs. Similarly, the sliding window remains the only used approach for input feed for non-sequential models, which is incorporated only by hybrid alignment models in the case of the sequential approach. This helps summarise that the applicability of component methodology is heavily driven and dependent on the learning techniques used.
   
\vspace{-1em}
    \subsection{Datasets}
    Benchmark datasets play a crucial role in the training and testing of deep learning VSR models. As reported in Table~\ref{tab:deep_res}, Vimeo90k~\cite{vimeo} is the most commonly used dataset for training because it is the largest dataset in the VSR domain and consists of diverse real-world scenes. The details of widely used datasets are given in Table~\ref{dataset}. Vid4~\cite{vid4} is the most common dataset used for test purposes, although it does not contain diverse scenes nor large inter-frame motion~\cite{RBPN}. YUV25\footnote{YUV25 - https://media.xiph.org/video/derf/} and CDVL\footnote{CDVL - https://www.cdvl.org/} have been used in earlier works for training and evaluation purposes. In order to train and test VSR models that account for video sequences with large inter-frame motion,  the most challenging property of video for VSR models, the REDS~\cite{REDS} dataset has been used since 2019. Other datasets, such as Myanmar\footnote{Myanmar- https://www.harmonicinc.com//insights/blog/4k-in-context/}, Ultra-Video Group\footnote{Ultra-Video Group Test Sequences - http://ultravideo.cs.tut.fi/} and SPMCS~\cite{SPMCS}, used by various published works for test purposes, are also listed and described in Table~\ref{dataset}.
        \begin{table*}[ht]
        \vspace{-0.5em}
        \centering
        \caption{Benchmark Datasets Commonly Used for Video Super-Resolution in the Literature.}

        \begin{tabular}{|p{2cm}|p{3.1cm}|p{12cm}|}
        \hline
        {\bf Dataset} & {\bf Resolution} & {\bf Description}\\
        \hline 
        YUV25\footnotemark[1] & 386×288p & 25 videos in the uncompressed YUV4MPEG format.\\
        Myanmar\footnotemark[3] & 3840×2160p& 59 scenes, uncompressed 4K resolution. \\
        CDVL\footnotemark[2] & 1920 × 1080 & 100 RGB videos from The Consumer Digital Video Library.\\
        Vid4~\cite{vid4}  & Calendar-720x576p, Walk–720x480p, Foliage-720x480p, City–704x576p &	The sequences in this dataset have visual artifacts, very little inter-frame variation, and limited motion. Most notably, the dataset consists of only four video sequences – City, Walk, Calendar and Foliage. \\
        Vimeo90K~\cite{vimeo} & 448x256p & A septuplet dataset consisting of 91,701 7-frame sequences with a fixed resolution 448x256, extracted from 39K selected video clips from the Vimeo-90K website. This dataset is designed for video de-noising, de-blocking, and super-resolution. Widely used for training as it contains a large number of clips with varied motions. \\
        Ultra-Video Group\footnotemark[4] &  1920×1080p & 7 HD videos of approx. 5 sec in length \\
        SPMCS~\cite{SPMCS} & 540×960p &975 sequences from commercial videos shot with high-end cameras and containing both natural-world and urban scenes with rich detail. Each sequence contains 31 frames. 945 sequences are randomly chosen as training data, and the rest 30 sequences are for validation and testing.\\
        REDS~\cite{REDS} & 1280x720& 240 video sequences with 100 frames in each, widely used for training and testing since 2019. Primarily includes clips with large inter-frame motions.\\
        \hline
        \end{tabular}
        \label{dataset}
        \vspace{-2em}
        \end{table*}
\vspace{-1em}
\subsection{Loss Functions}

There are various loss functions used to train VSR models in order to optimise learning. These functions can be broadly categorised into two groups: {\it pixel-based} and {\it perception-based.}

\subsubsection{Pixel-Based}

Pixel-based objectives aim to minimise the discrepancy between the generated HR frame and the corresponding ground truth (GT) frame. These objectives measure the pixel-wise differences between the generated HR and GT frames. Several commonly used reconstruction-based loss functions in VSR include:

\begin{itemize}
    \item {\it Mean Squared Error (MSE)}, also known as \textit{L2} loss, measures the average of the squared differences between the generated HR frame and the corresponding GT frame. MSE is a commonly used loss function in various image and video processing tasks. Minimising MSE encourages VSR models to generate HR frames that closely match the ground truth in terms of pixel values. However, MSE tends to emphasise large errors, which may lead to over-smoothed results, and MSE can produce blurry outputs. While MSE provides a straightforward and differentiable objective, optimising loss solely based on MSE can produce visually unsatisfactory results. It can, for instance, result in overly smooth outputs that lack fine detail and texture. MSE can be mathematically represented as:

    \vspace{-1em}
    \begin{small}
    \begin{equation}    
    L_{\mbox{\em MSE}} = \frac{1}{N} \sum_{i=0}^{N-1} (GT_i - HR_i)^2
    \label{eqn:mse}
    \end{equation}
    \end{small}
    
    where $GT_i$ represents the ground truth frame pixel at position $i$, $HR_i$ is the generated HR frame pixel at position $i$, and $N$ is the total number of pixels. 

    \item {\it Mean Absolute Error (MAE)}, also known as \textit{L1} loss, calculates the average of the absolute differences between the generated HR frame and the GT frame. Unlike MSE, MAE is less sensitive to outliers, making it more robust to extreme errors. By minimising MAE, the model is encouraged to produce HR frames that have accurate pixel-wise correspondence to the ground truth. MAE loss helps reduce the impact of outliers, which can be beneficial for handling noisy or corrupt data. MAE often leads to sharper and more detailed outputs compared to MSE loss. However, MAE may still produce over-smoothed results. MAE is expressed as:

    \vspace{-0.5em}
    \begin{small}
      \begin{equation}
    L_{\mbox{\it MAE}} = \frac{1}{N} \sum_{i=0}^{N-1} |GT_i - HR_i|
    \end{equation}
    \end{small} 
    
    \item {\it Smooth-\textit{L1} Loss}: Also known as Huber loss, combines the benefits of both \textit{L1} and \textit{L2} losses. Smooth-\textit{L1} Loss behaves in the same way as \textit{L2} loss when the pixel-wise differences are small (less sensitive to outliers) and as \textit{L}1 loss when the pixel-wise differences are large. The transition is controlled by a hyperparameter $\beta$. Smooth-\textit{L1} loss is advantageous because it suppresses the impact of outliers while still penalising large errors. Smooth-L1 loss can help strike a balance between accuracy and robustness. It tends to produce visually pleasing results with sharper details and fewer artifacts than MSE or MAE. The mathematical representation is: 

      \vspace{-0.5em}
      \begin{small}
          \begin{equation}
        \begin{aligned}
            L_{\mbox{\it Smooth-L1}} =
            \begin{cases}
               \frac{1}{N} \sum_{i=0}^{N-1} 0.5 (GT_i - HR_i)^2 \beta,\text{if } | GT - HR | < \beta \\
            \frac{1}{N} \sum_{i=0}^{N-1}|GT_i - HR_i| - 0.5\beta , \text{otherwise}
            \end{cases}
        \end{aligned}
        \end{equation}
      \end{small}

    where $\beta$ is a hyperparameter that controls the point where the loss transitions between \textit{L1} and \textit{L2} behaviour.

    \item {\it Charbonnier Loss}: Charbonnier Loss, also known as Charbonnier penalty (aka pseudo-Huber loss), is a differentiable and smooth approximation of \textit{L2} loss while having similar robustness properties as the \textit{L1} loss. By using the square root, Charbonnier loss reduces the weight of large errors and focuses more on smaller errors. This property makes it less sensitive to outliers and, therefore, more robust. Charbonnier loss can be an effective alternative to MSE and MAE, especially when dealing with noisy data or artifacts in the ground truth. It encourages the model to generate HR frames that preserve important visual details while suppressing the impact of noisy or inconsistent data and is defined as:

    \vspace{-0.5em}
    \begin{small}
    \begin{equation}
       L_{\mbox{\it Charbonnier}} = \frac{1}{N} \sum_{i=0}^{N-1} \sqrt{(GT_i - HR_i)^2 + \epsilon^2}
    \end{equation}
    \end{small}
    
    where $\epsilon$ is a small positive constant.

    \item {\it Adversarial Loss}: Adversarial loss employs a discriminator network to distinguish between the generated HR frame and real HR frames. It encourages the generated frames to be indistinguishable from the real HR frames. By training the model in an adversarial setting, it learns to capture finer details and textures in the generated HR frames, leading to more natural-looking outputs. Denoting the discriminator's output (probability of the frame being real) as $D(GT)$, for the ground truth HR frame and $D(HR)$, for the generated HR frame. Adversarial loss can subsequently be formulated as follows:

    \vspace{-0.5em}
    \begin{small}
    \begin{equation}
     L_{\mbox{\it Adversarial}} = \frac{1}{N} \sum_{i=0}^{N-1} \left[\log D(GT_i) + \log(1 - D(HR_i))\right]
    \end{equation}
    \end{small}
    
    In this adversarial loss formulation, the generator tries to maximise $\log(1 - D(HR_i))$, aiming to produce HR frames that are as close as possible to the pixel-wise representation of the discriminator.  On the other hand, the discriminator tries to maximise $\log D(GT_i)$ and $\log(1 - D(HR_i))$ to improve its ability to distinguish between real and generated frames.
 
    \item {\it Weighted Loss}: Weighted loss balances the contributions of different loss functions during training. By assigning weights to each loss, the model can focus on optimising specific aspects of the generated HR frames and customise the trade-off between different objectives. For example, a combination of pixel-wise loss (e.g., MSE) and perceptual loss can be used to achieve a balance between reconstruction accuracy and visual quality. Multiple loss functions are combined using weighted coefficients:

    \vspace{-0.5em}
    \begin{small}
        \begin{equation}
      L_{\mbox{\it Weighted}} = w_1 L_1 + w_2 L_2 + \ldots + w_n L_n 
   \end{equation}
    \end{small}
   
   where $L_j$ represents the $j^{th}$ loss function, and $w_j$ is the weight assigned to that loss function. The weight to be assigned is determined based on the relative importance of individual loss terms using manual tuning based on domain expertise or empirical validation of the relative impact of the individual components during training.

\end{itemize}
\subsubsection{Perception-Based}

Perception-based objectives focus on the perceived quality of the generated HR frame rather than pixel-wise differences. These objectives employ pre-trained networks, such as VGGNet~\cite{simonyan2014very} and AlexNet~\cite{krizhevsky2017imagenet}, to extract features from the GT and generated  HR frames, measuring the similarity between these features. The most commonly used perception objective-based loss function in VSR is perceptual loss. Perceptual loss utilises a pre-trained network to extract high-level features from the GT and generated HR frames. By comparing these features, the model is guided to minimise the difference in high-level representations rather than via pixel-wise error analysis. Perceptual loss aims to improve the visual quality of the generated HR frames by focusing on higher-level visual features. It encourages the model to capture the overall structure and content of the frames, resulting in outputs that are visually appealing to human observers. Perceptual loss is particularly effective in reducing the over-smoothing issue often encountered when relying solely on pixel-wise loss functions. Perceptual loss measures the similarity between these features. Mathematically, it can be expressed as:

    \begin{small}
        \begin{equation}
    L_{\mbox{\it Perceptual}} = \frac{1}{M} \sum_{k=0}^{M-1} (F(GT_k) - F(HR_k))^2
    \end{equation}    
    \end{small}
   
    where $F(GT_k)$ and $F(HR_k)$ are the features extracted from the ground truth frame and the generated HR frame, respectively, and $M$ is the total number of features.

   \vspace{-1em}
\subsection{Advanced Training and Optimisation}

To improve the performance and efficiency of VSR models, several advanced training and optimisation techniques are employed with the aim of enhancing the learning process and the generalisation capabilities of the VSR models. Some commonly used techniques in VSR include:

\subsubsection{Transfer Learning}

Transfer learning is a technique that leverages pre-trained model weights from other domains and adapts them to the VSR task. By utilising the knowledge learned from related tasks or domains, transfer learning allows VSR models to benefit from the representational power of pre-trained models and enhance their performance with even limited training data.
Transfer learning has been used for several purposes in the VSR literature. Initially, it was used to transfer the learning of a single image super-resolution model to a video super-resolution model as a base initialisation of the learning process~\cite{7444187}. With proven advances, transfer learning has also been used to fine-tune deep learning-based optical flow estimation methods such as SpyNet\cite{spynet}, in the particular context of video super-resolution task~\cite{BasicVSR, baniya2023online}. More recently, transfer learning has been used to adapt the super-resolution task from conventional 2D videos to the new domain of {360\textdegree} video super-resolution~\cite{baniya2023omnidirectional}, where the availability of the training dataset is limited. In each instance, transfer learning has proven to be an effective tool in extending the learning and performance ability of VSR models.

\subsubsection{Escaping Local Minima}

Deep learning VSR models are typically trained using gradient descent optimisation, which can sometimes get stuck in local minima, leading to suboptimal solutions. To address this issue, various techniques have been proposed to escape local minima and find better global minima. One such technique is to use a well-known variant of gradient descent called stochastic gradient descent (SGD). SGD introduces noise in the gradients and helps better explore the optimisation landscape. Another technique is to use a variant of SGD called Adam optimisation~\cite{adam}, which adapts the learning rate for each parameter and helps prevent getting stuck in sub-optimal local minima. The Adam optimiser remains the most commonly used optimisation technique in recent VSR literature. Recently, metaheuristic optimisation using learning rate reset has also proven to help expand VSR model learning ability despite premature training saturation~\cite{baniya2023online} while still improving inference results confirming no over-fitting. These approaches are guiding new directions in VSR research with an emphasis on advanced training and optimisation efforts. 

\subsubsection{Knowledge Distillation}

Knowledge distillation is a technique that allows the transfer of knowledge learned by a complex and larger model (the teacher) to a simpler and smaller model (the student). In the context of VSR, knowledge distillation was first explored by Xiao \textit{et al.}~\cite{xiao2021space} where the complex EDVR~\cite{EDVR} model was used to train a shallower student model for space and time distillation. Space distillation was done to train the student model to produce the teacher-like attention map, while time distillation was performed to help the student model learn weights for Convolution LSTM memory units from the teacher network. Adopting this technique, it was proven that multiple existing shallower networks were able to produce better results compared to conventional training. Several other works have since proven the effectiveness of knowledge distillation in VSR for the design of lightweight real-time VSR models~\cite{10041747}, to compress the model size of recurrent network~\cite{liu2021efficient} and to optimise the size of MEMC-based super-resolution model~\cite{lee2023knowledge}.

\subsubsection{Model Compression}

Model compression techniques aim to reduce the computational complexity and memory requirements of VSR models while preserving their performance. Techniques such as pruning, quantisation, and low-rank approximation can be applied to compress the model size and make it more suitable for deployment on resource-constrained devices or in real-time applications. Although earlier discussed, knowledge distillation techniques have also been used to train a compressed VSR model, the discussion here is focused on the methodology of compressing a trained network. In order to prune redundant filters, a recent work~\cite{Xia_2023_CVPR} proposed structured pruning schemes for residual blocks, recurrent networks, and upsampling networks. It was therein proven the application of pruning on BasicVSR~\cite{BasicVSR} resulted in up to $\times 4$ optimisation of efficiency while maintaining competitive qualitative and quantitative results. Agrahari Baniya et al.~\cite{baniya2023online} also applied the pruning technique to optimise the inference efficiency by removing the explicit alignment component from a trained model and instead retaining the alignment knowledge in convolution layers. This approach significantly improves model efficiency during inference enabling the lighter version to be used in an online application context with minimal performance degradation. 

Real-time resource-constrained applications are becoming a normative use case for VSR with the forecast growth of hand-held and edge devices. Advanced training and optimisation techniques are therefore the key enablers of VSR model development for these requirements~\cite{ignatov2021real}.

\vspace{-1em}
\section{Evaluation}

\subsection{Quality}
Quality evaluation in video super-resolution involves assessing the performance of the generated high-resolution (HR) video frames in reference to the ground truth (GT) frames. The assessment of quality is either subjective or objective. Predominantly, objective assessment methods based on pixel comparisons are used in the VSR domain. While the objective metrics help quantify the quality of produced HR frames and enable an analysis of the impact of Quality of Service (QoS), they often do not align with human perception of VSR quality. Most video-based multimedia is consumed through human perception, and thus there is a growing interest in evaluating VSR model performance in terms of perceptual quality. 
\rev{
\subsubsection{Peak Signal-to-Noise Ratio (PSNR)} PSNR is a widely used objective quality assessment metric in image and video processing tasks. PSNR measures the quality of the reconstructed video frame by comparing each co-located pixel with the original GT frame. PSNR is calculated as the ratio of the maximum possible pixel value to the Mean Squared Error (MSE) between the HR and GT frames. Higher PSNR values indicate better reconstruction quality since its value signifies less distortion and a closer resemblance to the GT. However, PSNR has limitations in capturing perceptual differences and may not always correlate well with human visual perception. PSNR is calculated as:

\vspace{-0.5em}
    \begin{small}
        \begin{equation}
        \text{PSNR}_{(GT,HR)} = 10 \log_{10} \left(\frac{MAX^2}{MSE_{GT,HR}}\right)
    \end{equation}
    \end{small}
    
    Where $MAX$ is the maximum possible pixel value of the image (usually 255 for an 8-bit frame), and $MSE_{GT,HR}$ is the Mean Squared Error between the GT and the reconstructed HR frames computed using Eqn.~\ref{eqn:mse}.

\subsubsection{Structural Similarity Index Measure (SSIM)} SSIM is another objective image quality metric that evaluates the structural similarity between the HR and GT videos by considering luminance, contrast, and structural information. The SSIM index ranges between -1 and 1, with 1 indicating perfect similarity. SSIM accounts for both global and local spatial structural information and is known to correlate better with human perception compared to PSNR. SSIM can be computed as:

\vspace{-1em}
    \begin{small}
       \begin{equation}
        \text{SSIM}_{(GT,HR)} = \frac{(2\mu_{GT}\mu_{HR} + c_1)(2\sigma_{GT,HR} + c_2)}{(\mu_{GT}^2 + \mu_{HR}^2 + c_1)(\sigma_{GT}^2 + \sigma_{HR}^2 + c_2)}
    \end{equation} 
    \end{small}
    
    Where $\mu_{GT}$ and $\mu_{HR}$ are the means of the image $GT$ and $HR$, $\sigma_{GT}^2$ and $\sigma_{HR}^2$ are the variances of the image $GT$ and $HR$, and $\sigma_{GT,HR}$ is the covariance of the image $GT$ and $HR$. The constants $c_1$ and $c_2$ are used to stabilise the division with a weak denominator.
    
\subsubsection{ Natural Image Quality Evaluator (NIQE)} NIQE measures the naturalness of a frame where no reference is provided~\cite{6353522}. The degree of naturalness is computed by analysing image statistics such as luminance, contrast and edge information. Higher NIQE scores indicate higher perceived naturalness in the frame and are useful in assessing the perceived quality of super-resolved frames, as it considers characteristics important for human visual perception beyond traditional pixel-based metrics like PSNR and SSIM. 

\subsubsection{ Learned Perceptual Image Patch Similarity (LPIPS)} is a perceptual metric that measures the perceptual distance between two frames using a neural network like VGG16. LPIPS quantifies the visual dissimilarity between the HR and GT frames, taking into account high-level features extracted from the VGG model, which has learnt feature extraction from a large database of images. LPIPS is designed to capture human perceptual judgments and is more in line with human visual perception compared to PSNR and SSIM. 

\vspace{-0.5em}
    \begin{small}
       \begin{equation}
        \text{LPIPS}_{(GT,HR)} = \|VGG_\theta(GT) - VGG_\theta(HR)\|_2
    \end{equation} 
    \end{small}
    
    Where $VGG_\theta$ is the neural network, $GT$ and $HR$ are the frames being compared, and the norm is the $L2$-norm. Unlike the traditional quality metrics such as PSNR and SSIM, LPIPS is a trained metric using a neural network and has only recently started being used for evaluation. It is still not a common assessment metric in the VSR literature; however, it is commonly used to evaluate models in competitions like NTIRE~\cite{li2023ntire}. 

\subsubsection{Temporal Consistency} Although not commonly used, it is an important perceptual aspect in VSR to evaluate the temporal coherence between consecutive frames in the generated HR video sequence. Temporal consistency metrics such as flow warping error~\cite{lai2018learning} assess the maintenance of continuity and smoothness over time, ensuring that there are no noticeable artifacts or inconsistencies between frames. High temporal consistency indicates that the VSR model effectively preserves the motion dynamics and temporal details present in the GT, leading to visually pleasing results.}

\vspace{-1em}
 \subsection{Efficiency}

Efficiency is a critical factor to consider when evaluating video super-resolution models. Traditionally, the key focus on VSR model evaluation was limited to quality assessment. However, with the growing computational demand of neural networks and constraints around resource availability, efficiency evaluation during inference has become a standard evaluation method. This section explores key efficiency metrics, including model size, inference time, frames per second (FPS), and floating-point operations (FLOPs).

\subsubsection{Model Size}
The model size denotes the memory requirements for storing parameters and architecture of a video super-resolution model. Model size influences the feasibility of deploying the model on resource-constrained devices or in scenarios where storage capacity is limited. Smaller model sizes are advantageous, as they enable efficient utilisation of storage resources. The size of a deep learning video super-resolution model can be computed by considering the number of parameters it contains. Denoting the total number of layers in the model as $P$ and the number of parameters in each layer as $L_l$, where $l$ represents the layer index. The total number of parameters in the model, $T$, can then be calculated by summing the parameters across all layers:

\vspace{-0.5em}
\begin{small}
    \begin{equation}
\label{eqn:model_size}
T = \sum_{l=0}^{P-1} L_l
\end{equation}
\end{small}

Each layer may have different types of parameters, such as weights, biases, or other learnable parameters specific to the layered architecture. Therefore, $L_l$ can vary depending on the layer type and its configuration.
For example, in a fully connected (dense) layer, $L_l$ would be the product of the number of input units, $n_{\text{in}}$, and the number of output units, $n_{\text{out}}$, along with any additional parameters such as biases:

\vspace{-0.5em}
\begin{small}
    \begin{equation}
L_l = n_{\text{in}} \times n_{\text{out}} + n_{\text{out}}
\end{equation}
\end{small}

Similarly, in a convolutional layer, the number of parameters depends on the filter size, the number of input channels, the number of output channels, and any biases:

\vspace{-0.5em}
\begin{small}
    \begin{equation}
\begin{aligned}
L_l = (\text{filter\_width} \times \text{filter\_height} \times \text{input\_channels} + 1)\times \text{output\_channels}
\end{aligned}
\end{equation}
\end{small}

By summing the parameters across all layers, as shown in eqn.~(\ref{eqn:model_size}), the total size of the deep learning video super-resolution model is determined.

\subsubsection{Inference Time}
Inference time ($t_{\text{inf}}$) represents the duration taken by a model to process a single video frame and generate the corresponding super-resolved output. Minimising inference time is important, particularly for real-time applications, where prompt processing and smooth playback are essential. The inference time of a deep learning video super-resolution model can be computed by considering the number of frames, the model architecture, and the hardware specifications. Denoting the total number of frames in the video as $F$, and the time taken to process one frame as $t_{\text{frame}}$. Additionally,  assume that the model processes each frame independently without any temporal dependencies. The total inference time, $T_{\text{total}}$, can then be calculated by multiplying the number of frames by the time taken to process one frame:

\vspace{-0.5em}
\begin{small}
    \begin{equation}
t_{\text{total}} = F \times t_{\text{frame}}
\label{eqn:totaltime}
\end{equation}
\end{small}

The time taken to process one frame, $t_{\text{frame}}$, depends on various factors such as the model architecture, the input size of each frame, and the hardware specifications. The time taken can be further decomposed into the time taken for computation, memory access, and any additional overhead:

\vspace{-0.5em}
\begin{small}
    \begin{equation}
t_{\text{frame}} = t_{\text{computation}} + t_{\text{memory}} + t_{\text{overhead}}
\label{eqn:time}
\end{equation}
\end{small}

The computation time, $t_{\text{computation}}$, represents the time taken by the model to perform the necessary computations for each frame. This depends on the complexity of the model architecture and the number of operations required. The memory access time, $t_{\text{memory}}$, accounts for the time taken to load input frames, intermediate results, and store output frames in memory. This depends on the memory bandwidth and access patterns. The overhead time, $t_{\text{overhead}}$, includes any additional time required for data preprocessing, postprocessing, or any other operations specific to the model or the implementation. By summing the computation, memory access, and overhead times as shown in eqn.~(\ref{eqn:time}), the time taken to process one frame can be determined. Multiplying it with the number of frames, as shown in eqn.~(\ref{eqn:totaltime}), gives the total inference time of the video super-resolution model.

\subsubsection{Frames per Second (FPS)}
The Frames per Second (FPS) metric measures the number of video frames that a model can process per second. Higher FPS values indicate faster processing capability and improved real-time performance. Models with high FPS can effectively handle videos with higher frame rates, ensuring seamless and continuous output without compromising quality. FPS can be computed as the inverse of the inference time per frame. Given that the inference time for a frame is in seconds, FPS is:

\vspace{-0.5em}
\begin{small}
    \begin{equation}
    FPS = \lfloor1/t_{\text{frame}}\rfloor
\end{equation}
\end{small}

where $\lfloor$  $\rfloor$ represents the rounded value for whole number.

\subsubsection{Floating Point Operations (FLOPs)}
Floating Point Operations (FLOPs) provide an estimation of the computational complexity of a video super-resolution model. FLOPs quantify the number of floating-point arithmetic operations required for each inference. Evaluating FLOPs is essential, as lower FLOP counts generally indicate more computationally efficient models. Such models are advantageous for deployment on devices with limited computational power or in scenarios where energy consumption is a concern. The computation of $\text{FLOPs}$ can be estimated using equations based on the model architecture and the number of operations performed by each layer. These equations depend on the specific architecture and operations involved and can be derived accordingly.

        \vspace{-1em}
\section{Applications}
\subsection{Spatial Sciences and Geography}

\rev{Video super-resolution techniques have proven to be invaluable in the field of spatial sciences and geography. By enhancing the spatial resolution of video content, finer details and identifying patterns can be extracted, enabling accurate analysis and understanding of various geographical phenomena. An application of video super-resolution in this domain is for satellite imagery that captures large amounts of video data from space, but the resolution is often limited due to complex imaging conditions and unknown degradation process~\cite{xiao2021satellite,jiang2018progressively}. VSR can enhance the resolution of satellite videos allowing for more precise monitoring of land cover changes, urban growth, and environmental phenomena such as deforestation or glacier melting~\cite{liu2020satellite,xiao2023deep}. Additionally, STVSR can also be used to obtain enhanced motion dynamics for extreme events observation while further improving spatial super-resolution~\cite{xiao2022space}. However, challenges persist in variation of scale across the satellite imagery and scarce motion in long-time series satellite video frames.} Additionally, video super-resolution techniques find application in geographic surveillance systems. Surveillance cameras installed in urban or natural environments often capture low-resolution video footage, which can hinder the identification of important details. Super-resolution algorithms can enhance the resolution of these videos, enabling more accurate object detection, tracking, and recognition. This is particularly useful in scenarios such as crowd monitoring or traffic analysis.

\vspace{-1em}
\subsection{Streaming Entertainment}

Video streaming in the entertainment industry has greatly benefited from video super-resolution techniques, enhancing the viewing experience for audiences. With the increasing demand for high-quality content on Video streaming platforms such as Netflix, Amazon Prime, and YouTube, video super-resolution plays a vital role in upscaling lower-resolution videos to meet the expectations of its viewers.\rev{Streaming services can enhance the resolution of videos in real-time or during post-production~\cite{zhang2020improving} allowing user content to have improved visual quality providing an improved viewing experience~\cite{zhang2021video}. Super-resolution techniques also help preserve the integrity of older or archived content by upscaling it to modern display standards. Similarly, specialised super-resolution networks, such as those tailored for face super-resolution employing purpose-built network architectures~\cite{yu2021efficient} can further assist in improving the perceived QoE of the videos streamed.} 

\vspace{-1em}
\subsection{Extended Reality}

Extended reality (XR), which encompasses virtual reality (VR), augmented reality (AR), and mixed reality (MR), relies heavily on high-resolution video content for creating realistic and immersive experiences. Video super-resolution techniques play a crucial role in enhancing the visual quality of XR content. In VR applications, super-resolution can be used to improve the resolution and clarity of {360\textdegree} videos, creating more immersive virtual environments. Higher-resolution videos reduce pixelation and blurriness, providing users with a more detailed and realistic VR experience~\cite{liu2020single,baniya2023omnidirectional}. In AR and MR applications, video super-resolution can enhance the quality of real-time video feeds from cameras, allowing virtual objects or information to be seamlessly integrated into the user's view~\cite{mora2023video}. By upscaling the video feed, super-resolution ensures that the virtual elements blend more smoothly with the real environment, enhancing the overall visual fidelity and user experience.

\vspace{-1em}
\subsection{Agriculture}

Video super-resolution has significant applications in the field of agriculture, enabling more detailed and accurate analysis of crop health, vegetation patterns, and agricultural processes. By enhancing the resolution of aerial or ground-based videos, researchers and farmers can gain valuable insights into crop management and precision agriculture. In aerial imaging, drones equipped with cameras capture videos of agricultural fields. However, due to altitude, distance, and limitations of drone cameras, the captured footage may have a lower resolution. Video super-resolution can help overcome this limitation by enhancing the resolution, allowing for more precise monitoring of crop growth, disease detection, and assessment of irrigation needs. Similarly, in ground-based imaging, video super-resolution can enhance the resolution of videos captured by fixed or moving cameras in the field~\cite{baniya2023current}. This aids in monitoring plant growth, detecting anomalies, and optimising farming practices. The higher resolution enables the identification of small-scale variations in crop health, pest infestations, or nutrient deficiencies, facilitating timely interventions for improved yield and sustainability.

\vspace{-1em}
\subsection{Health}

Video super-resolution techniques find many applications in the healthcare industry, enabling enhanced visualization and analysis of medical imaging videos~\cite{ren2021medical}. Medical professionals often rely on high-resolution videos for accurate diagnosis, treatment planning, and surgical interventions. In medical imaging, such as endoscopy or laparoscopy, video super-resolution can improve the visibility of fine structures, allowing physicians to detect abnormalities or lesions that may have been difficult to observe in lower-resolution video signals. This assists in the early detection of diseases, precise surgical guidance, and minimally invasive procedures. Furthermore, in telemedicine and remote patient monitoring, video super-resolution plays a vital role in transmitting high-quality video feeds over limited bandwidth connections. By enhancing the resolution of real-time video streams, healthcare providers can assess patients' conditions more accurately, enabling remote diagnosis, consultation, and continuous monitoring.

\vspace{-1em}
\subsection{Transport and Road}

\rev{Video super-resolution techniques have practical applications in the domain of transport and road infrastructure, contributing to improved safety, traffic management, and surveillance systems. In traffic monitoring, surveillance cameras installed on roads capture video footage of vehicles and pedestrians. However, these videos often suffer from low resolution, making it challenging to identify license plates, read signage, or detect fine details for incident analysis~\cite{guo2023video}. In autonomous vehicles and advanced driver-assistance systems, (ADAS)~\cite{daithankar2021adas}, high-resolution video inputs are crucial for accurate object detection, lane tracking, and collision avoidance. Super-resolution techniques can enhance the resolution of camera feeds, providing a clearer view of the surroundings and improving the performance and safety of autonomous systems. Additionally, video super-resolution is valuable in infrastructure monitoring, where cameras are deployed to assess the condition of roads, bridges, and tunnels. Enhanced resolution videos help identify structural defects, cracks, or signs of deterioration, allowing for timely repairs and proactive maintenance and safety strategies.}
 \vspace{-1.2em}       
\section{Challenges and Trends}
 \vspace{-0.75em}     
\subsection{Online VSR}
 \vspace{-0.75em}     
A growing trend in the current VSR literature, as seen in Tables~\ref{tab:summary} and ~\ref{tab:deep_res}, is to model the temporal correlation across the time domain effectively. To do so, non-sequential models tend to extend the size of the sliding window to increase temporal context from a wider radius~\cite{RBPN}, and sequential models tend to use bidirectional recurrent models which expect all the video frames to be available during inference~\cite{BasicVSR}. Both approaches are limited in their ability for online usage such as streaming, video conference, etc., where limited frames are available for each timestamp. One ideal solution is to use unidirectional RNNs, which have proven to be effective while using single frame input per timestamp. However, the lack of information for earlier frames results in compromised performance for the earlier frames in the video series~\cite{RRN}. 

\vspace{-1.2em}
\subsection{\rev{Real VSR}}
\rev{While deep learning VSR models have proven to be effective in restoring high-frequency details from low-resolution video frames downgraded synthetically, their extension to real VSR also known as blind VSR, where the degradation is unknown, remains challenging. Real VSR or blind VSR should encompass diverse degradations, and a single dataset may not represent this diversity well. To address these challenges, modifications involving a cleaning task as a pre-processing step~\cite{chan2022investigating}, latent feature transformation~\cite{pan2021deep}, and diverse kernels combined into VSR for input adoption~\cite{lee2021dynavsr}, have been proposed in the literature. Additionally, a decomposition-based learning scheme has been proposed, where LR-HR videos are converted into YUV colour space and the luminance channel is decomposed into a Laplacian pyramid, followed by applying different loss functions to different components, resulting in VSR models with improved performance under real-world settings~\cite{yang2021real}. Approaches to synthetically create larger datasets with randomised degradation in generating LR video sequences have also been used to train end-to-end VSR models~\cite{Jeelani_2023_CVPR}. Unfortunately, all methodologies in deep learning VSR attempt to build a single model solution to generalise the diverse degradations in blind VSR tasks. Alternative methodologies are for diverse degradations in real video inputs without attempting to train a single model to super-resolve all diverse blind video input types.}

\vspace{-1em}
\subsection{Resource Constrained VSR}

Resource-constrained VSR is a significant challenge when deploying super-resolution models on devices with limited computational resources. Real-time applications and on-device processing require efficient algorithms that can produce high-quality results without overwhelming the limited hardware~\cite{ignatov2021real}. To address resource constraints, lightweight neural network architectures, those that strike a balance between model complexity and performance, such as MobileNet~\cite{howard2017mobilenets} or EfficientNet~\cite{tan2019efficientnet}, have been proposed. These models can significantly reduce the number of parameters and computational costs while maintaining reasonable super-resolution quality. Additionally, techniques such as knowledge distillation~\cite{hinton2015distilling} and network pruning~\cite{baniya2023online} have been employed to reduce the model size further and improve inference speed without significant loss in VSR accuracy. However, the need for higher resolution and faster inference is growing as technology and consumer behaviours rapidly evolve, resulting need to further optimise VSR \revb{computational} efficiency. 

\vspace{-1em}
\subsection{Large and Scalable Upscaling}

Large upscaling in VSR involves increasing the resolution of low-resolution videos by a significant factor, typically beyond $\times 4$. Handling large upscaling factors presents unique challenges, as the models need to generate high-frequency details that are significantly missing in the low-resolution input.VSRnet~\cite{7444187} demonstrated the performance degradation as the scaling factor increased from $\times 2$ to $\times 3$ to $\times 4$. \rev{Considering $\times 4$ super-resolution as the most representative task, many of the recent VSR models only train and evaluate the model for $\times 4$ super-resolution.} The ability of current VSR models to super-resolve beyond that factor and their corresponding solutions remain mostly unexplored in the literature. On the other hand, scalable upscaling expects VSR models to generate a wide range of upscaling factors effectively. For real-world applications, different videos may require various upscaling factors based on their original resolution and the desired output resolution. Fixed-factor VSR models may not be optimal in such scenarios, as they may either overspend computational resources on small upscaling factors or fail to produce satisfactory results for larger factors. A prominent trend in scalable upscaling is the use of progressive upscaling techniques, where the models generate intermediate resolutions before reaching the final high-resolution output. Progressive upscaling allows the model to adapt to different upscaling factors, producing improved results across a wide range of resolutions. For instance, models like RBPN~\cite{RBPN} have demonstrated success in scalable upscaling by progressively increasing the spatial dimensions of the generated frames, thereby catering to different upscaling requirements efficiently. In real-world uses, VSR models need to attain enhanced adaptability as well as performance with varying scaling factors.

\vspace{-1.3em}
\subsection{Multi-objective VSR}
Multi-objective VSR refers to the task of simultaneously addressing multiple objectives or criteria in the video super-resolution process. These objectives may include enhancing resolution, improving visual quality, reducing artifacts, and preserving important visual details. Traditionally, VSR models have focused on a single objective, such as maximising a pixel-based accuracy, such as peak signal-to-noise ratio (PSNR) or structural similarity (SSIM) index, to measure the similarity between the super-resolved and ground truth frames. In multi-objective VSR, the goal is to strike a balance between different objectives, considering that they may sometimes conflict with each other. For example, increasing the resolution might introduce artifacts or result in over-smoothed textures, negatively affecting visual quality. To address this challenge, recent research has focused on developing VSR models that can handle multiple objectives simultaneously. One common approach in multi-objective VSR is to use advanced weighted loss functions that combine different metrics, such as a combination of pixel-based losses and perceptual loss based on deep features extracted from pre-trained neural networks such as VGG or ResNet. These combined loss functions allow the model to be optimised for both perceptual quality and objective measures simultaneously. Another trend in multi-objective VSR is to introduce regularisation terms, or additional components in the loss function, that explicitly encourage specific desirable characteristics. For example, to improve the sharpness of super-resolved frames, an edge-aware regularisation term can be included in the loss function to preserve fine details and avoid over-smoothing. Training multi-objective VSR models can be challenging since it involves finding the appropriate trade-offs between different objectives and fine-tuning model parameters accordingly. Moreover, multi-objective VSR may add complexity to the optimisation process, potentially leading to longer training times, difficult convergence and increased computational resources. 
\vspace{-1.3em}
\rev{
\subsection{Space-Time VSR}
Space-time VSR (STVSR) models aim to improve video quality across both spatial and temporal dimensions. Incorporating the increased dimensionality of the temporal interpolation directly into conventional VSR without introducing artifacts or distortions while maintaining computational efficiency is challenging. As such, research in STVSR is directed towards efficiently leveraging spatiotemporal correlations while addressing the challenges of simultaneously interpolating time and space details~\cite{1401907}. Recent advancements in STVSRs utilise deep learning techniques such as spatial-temporal transformers~\cite{geng2022rstt}, deformable convolution LSTMs~\cite{xiang2020zooming,xu2021temporal} and mutual learning through iterative up and downsampling~\cite{10177211} to jointly model spatial and temporal dependencies. Temporal profile along with spatial-temporal fusion has also been used to directly exploit the spatial-temporal correlation in the long-term temporal context without the need for motion compensation\cite{10.1145/3394171.3413667}. On the other hand, synthesising LR frames using flow maps and blending masks, and reusing upsampled versions of these for coarse estimation of HR intermediate frame followed by refinement using residual learning has also resulted in light-weight STVSR models~\cite{dutta2021efficient}. Despite these advancements, addressing issues related to computational complexity, scalability to high-resolution videos, and generalisation across diverse video content are among the key challenges in this field.}
\vspace{-1em}
\section{Conclusion}
A thorough overview and systematic categorisation of each VSR component and technology have been presented in this work contributing to a categorical taxonomy and summarisation of VSR practices with highlights of key methods and their implications across VSR stages. Despite the diverse spread of the tools and technologies used in VSR, the critical review introduced in this paper thoroughly summarises the preeminent choices. Furthermore, this study provides detailed guidelines for the deep learning technologies that are currently used in the literature and unpacks the substance of each technology, ranging from VSR model architecture to model training and evaluation. With additional discussion of the application of VSR models and the challenges and trends in the field, this study aims to foster a guide for VSR research with explainable technology selections and requirement-specific modelling. Overall, this study serves as a comprehensive resource for researchers, practitioners, and stakeholders in the field of VSR, facilitating practice-based decision-making and advancing the state-of-the-art in deep learning-based video super-resolution. \rev{There needs to be further analysis into the explainability of the combined impact of these components in relation to the results obtained, and this work is intended to set the stage for that.}
\vspace{-1em}
\balance

\bibliographystyle{IEEEtran}
\bibliography{cas-refs}

\end{document}